%% file: main.tex
\documentclass[12pt, english]{article}
\usepackage{amsmath,amsfonts}
\usepackage{algorithm}
\usepackage{algcompatible}
\usepackage{algpseudocode}
\algnewcommand\RETURN{\State \algorithmicreturn}
\algnewcommand\And{\textbf{and}}
\algnewcommand\Or{\textbf{or}}

\usepackage{array}
\usepackage{textcomp}
\usepackage{stfloats}
\usepackage{url}
\usepackage{verbatim}
\usepackage{graphicx}
\usepackage[dvipsnames]{xcolor}
\usepackage{subcaption}
\usepackage{multirow}
\usepackage{hyperref}
\usepackage{cleveref}
\usepackage{tabularx}
\usepackage{tabularray}
\usepackage{adjustbox}
\usepackage{nicematrix}
\usepackage{diagbox}
\usepackage{makecell}
\usepackage{boldline}
\usepackage{booktabs} 
\usepackage{biblatex}
\providecommand{\keywords}[1]{\textbf{\textit{Keywords ---}} #1}
\graphicspath{{./figures/}}

\begin{document}

\title{Learning QoE from Packet-Level Measurements in Encrypted Video Conferencing Traffic}

\author{Michael Sidorov, Ofer Hadar}


\maketitle

\input{sections/abstract}

\input{sections/introduction}

\input{sections/related_work}

\input{sections/methodology}

\input{sections/datasets}

\input{sections/experiments}

\input{sections/ablation}

\input{sections/results}

\input{sections/conclusions_and_future_work}

\printbibliography

\input{sections/appendix}

\end{document}

%% file: sections/abstract.tex
\begin{abstract}
The quality of the user experience has become one of the most important aspects in today’s world, as it directly influences individuals’ willingness to continue using or abandon a product or service. In this context, video conferencing applications (VCAs), which experienced widespread adoption following the COVID-19 pandemic, must deliver excellent performance to remain competitive in an increasingly crowded market. Although content providers (CPs) such as Zoom, WhatsApp, Telegram, and Google Meet can assess conversation quality by comparing transmitted and received data. The widespread use of end-to-end encryption in VCAs makes quality-of-experience (QoE) evaluation by internet service providers (ISPs) far more challenging. Since ISPs do not have access to the encrypted content, they must rely on passive measurements of unencrypted traffic characteristics on the data path. In this work, we present a simple yet effective QoE prediction framework based on an almost stock convolutional neural network (CNN) architecture that uses only the packet sizes extracted from the communication between two participants in a video conferencing (VC) call to predict two QoE metrics: BRISQUE and MOS. The proposed framework is simple, easy to implement, and does not require high-end computational resources, yet it provides superior prediction performance, as shown in our experiments on two custom datasets collected from WhatsApp and Zoom, which achieve substantial improvements over previous models for the QoE prediction task.
\end{abstract}

\keywords{Quality of Experience in Video, Encrypted Traffic, Deep Learning, Machine Learning, Video Conferencing}

%% file: sections/introduction.tex
\section{Introduction}
Predicting the Quality of Data (QoD) delivered to the end user is one of the central goals in modern communication systems, as it directly shapes users' Quality of Experience (QoE). QoE, in turn, strongly influences whether users continue using (or abandon) a specific Content Provider (CP) or Internet Service Provider (ISP).

The most direct way to assess QoD (and consequently QoE) is to compare the received content with the original data transmitted by the server, calculate a measure of similarity, and act to improve it. In practice, however, such end-to-end comparison is feasible primarily for CPs, since they control both ends of the delivery pipeline. For ISPs to obtain similar insight, they would need to intercept and analyze transferred packets, a practice historically known as Deep Packet Inspection (DPI), which was widely used during the 1990s and early 2000s.  

Since approximately 2010, the rapid growth of the internet and increased privacy concerns have made end-to-end encryption a fundamental requirement for most online communications. As a result, traditional DPI has become largely ineffective for ISPs \cite{naylor2014cost}, because encrypted traffic payloads cannot be interpreted without access to private keys.

Consequently, ISPs must primarily rely on passive measurements of network Quality of Service (QoS) indicators \cite{feamster2017and}. Their objective is to infer QoD from observed QoS metrics and ultimately predict users' QoE. This inference is inherently challenging due to the complex, highly non-linear relationship between QoS and QoE, which has motivated extensive research over the past decades.

Though it is natural to think of representative features for each traffic feature and manually match them to the corresponding QoE metric, for high-volume applications (such as video streaming), a data-driven approach is particularly natural. Accordingly, numerous machine learning (ML) and deep learning (DL) methods have been proposed to infer various QoE measurements (such as buffer state, jitter, or available bandwidth) under encrypted traffic constraints, achieving progressively improved accuracy.

Despite improvements in prediction accuracy, ML, and especially DL models, suffer from overfitting when the dataset is small or not diverse enough to represent all possible instances in the test set. 

In this study, we aim to bridge this gap by introducing qCNN, a simple, lightweight framework based on a convolutional neural network (CNN) architecture for QoE prediction from encrypted traffic. qCNN surpasses the performance of other models commonly used in time-series data on the QoE prediction task and provides an open-source solution for QoE prediction in encrypted traffic. We hope this contributes to wider accessibility and fosters further advancements in the field.
The main contributions of this paper are as follows:

\begin{itemize}
\item We propose a novel CNN-based framework for QoE prediction from passive low-level traffic measurements collected from encrypted communication channels, without relying on application-level information or payload inspection.

\item We investigate the stationarity properties of low-level traffic measurements and demonstrate that, for closed communication platforms such as WhatsApp and Zoom, traffic fingerprinting can be effectively leveraged for QoE prediction.

\item We introduce a learnable one-dimensional to two-dimensional traffic representation based on an embedding (EMBD) module and demonstrate that transforming traffic sequences into two-dimensional feature spaces improves the effectiveness of CNN-based feature extraction.

\item We present and publicly release two custom QoE datasets collected on WhatsApp and Zoom platforms. In particular, the Zoom dataset is annotated using Mean Opinion Score (MOS) labels, providing a rare source of subjective QoE measurements for video conferencing applications.

\end{itemize}

The remainder of this paper is organized as follows. \Cref{sec:related} surveys recent studies on QoE prediction, and highlights works using CNNs to improve performance across domains. \Cref{sec:dataset} analyzes the datasets used in our experiments. \Cref{sec:methodology} presents the proposed method, and \Cref{sec:experiments} describes the experiment setup employed during models' performance evaluation. \Cref{sec:results} reports the experimental results, and finally, \Cref{sec:conclusions} concludes the paper, discusses limitations, and outlines directions for future research.

%% file: sections/related_work.tex
\section{Related Work}
\label{sec:related}
The domain of QoE analysis has been extensively studied over the years. Bentaleb et al. \cite{bentaleb2019bandwidth} introduced a Recursive Least Squares (RLS) model to predict the instantaneous available network bandwidth. Their goal was to improve QoE by optimizing the data chunk size within the Dynamic Adaptive Streaming over HTTP (DASH) protocol. Through their method, the authors reported improved achieved bitrates and a reduction in stall events during video streaming sessions.

Numerous studies have proposed NN and DL–based methods for QoE prediction under the constraint of encrypted traffic. For example, Vega et al. \cite{torres2017unsupervised} introduced an unsupervised DL model based on a Restricted Boltzmann Machine (RBM) \cite{hinton2006reducing} for real-time prediction of various No-Reference (NR) QoE metrics in video streaming. Their model estimates indicators such as video jerkiness, image blockiness, and multiple blur and noise-related features. Although they report an accuracy exceeding 85\%, this performance is achieved only for a specific type of video content, limiting the generalization of their approach.

Raca et al. \cite{raca2020leveraging} explored two classical ML methods, viz. Support Vector Machines (SVM) and Random Forests (RF), alongside a DL approach using Long Short-Term Memory (LSTM) networks to predict throughput in cellular networks. Their experiments demonstrated that increasing the granularity of the input data improved performance, reflected in a reduced absolute error percentage between predicted and actual throughput.

A Reinforcement Learning (RL)–based framework was proposed by Mao et al. \cite{mao2017neural}, which computes three QoE metrics using predicted values of bitrate, available bandwidth, and buffer level in video sessions. Their results show that the RL model significantly outperforms traditional Adaptive Bitrate (ABR) mechanisms for QoE optimization.

Eswara et al. \cite{eswara2019streaming} proposed LSTM-QoE, a continuous Quality of Experience (QoE) prediction framework based on Long Short-Term Memory (LSTM) networks. The method operates on sequential data and is motivated by the observation that a user's QoE depends not only on the current video quality but also on previously experienced quality impairments and rebuffering events. Consequently, QoE is modeled as a nonlinear, dynamic, and non-Markovian process exhibiting both short- and long-term temporal dependencies. To capture these dependencies, the authors designed a multi-layer, multi-unit LSTM architecture that continuously predicts user QoE from a set of QoE-related features. The proposed model employs a cascade of LSTM layers, enabling it to learn complex temporal relationships and nonlinear patterns governing QoE evolution over time.

Although LSTM-QoE achieved state-of-the-art performance on several public QoE datasets, it has several limitations. Most notably, the model relies on application-level features, including Short Time Subjective Quality (STSQ), Playback Indicator (PI), and Time Since Last Rebuffering (TR). Such information is typically unavailable under encrypted traffic conditions, limiting the applicability of the method to passive network-side QoE estimation. Furthermore, the sequential nature of LSTM networks introduces higher computational complexity and inference latency compared to convolutional architectures, which may hinder deployment in real-time, large-scale monitoring systems.

Recently, convolutional neural network (CNN)-based models have been introduced for QoE prediction. For example, Bai et al. \cite{bai2018empirical} introduced the Temporal Convolutional Network (TCN), a convolutional architecture for sequence modeling proposed as an alternative to recurrent neural networks (RNNs), including LSTM and Gated Recurrent Unit (GRU) models. TCN combines the ability of convolutional neural networks (CNNs) to efficiently extract local and hierarchical features with causal convolutions that preserve the temporal ordering of sequential data. By employing dilated causal convolutions and residual connections, TCN is capable of modeling long-range temporal dependencies while maintaining computational efficiency through parallel processing.
In addition, TCN performance strongly depends on the selection of architectural hyperparameters, including the receptive field size, dilation factors, kernel size, and network depth. Since the optimal configuration is often domain-dependent, significant tuning may be required when transferring the model between applications. Furthermore, an insufficient receptive field may prevent the model from capturing long-term dependencies, whereas an excessively large receptive field may increase computational complexity and the risk of overfitting, potentially limiting the robustness and generalizability of the approach.

Duc et al. \cite{duc2020convolutional} proposed a one-dimensional CNN architecture incorporating causal and dilated convolutional layers. The authors demonstrated that causal 1D convolutions can effectively capture temporal dependencies in sequential QoE data while maintaining computational efficiency. Their results showed that the proposed approach outperformed several models specifically designed for sequence modeling, including Long Short-Term Memory (LSTM)-based architectures.

Nevertheless, the proposed method exhibits several limitations. First, it relies on application-level features, such as Spatio-Temporal Reduced Reference Entropic Differencing (STRRED), playback indicators, the number of rebuffering events, and the time elapsed since the last impairment, all of which are unavailable when traffic is encrypted and application-layer information cannot be accessed. Second, the model operates on a relatively short temporal context, using a receptive field of at most 20 time steps. While this design choice is motivated by the recency effect discussed by the authors, it may not generalize to scenarios in which QoE is influenced by longer-term temporal dependencies or cumulative user experiences.

\subsubsection{1D to 2D transformation}
\label{subsub:1d_2d}
A conceptually related approach was proposed by Shapira and Shavitt \cite{shapira2019flowpic}, who introduced FlowPic, a framework that transforms packet-level traffic traces into two-dimensional histograms and subsequently applies convolutional neural networks for encrypted traffic classification. Their results demonstrated that representing traffic as an image allows CNNs to achieve excellent classification performance without relying on handcrafted statistical features. However, the FlowPic representation is manually designed and relies on predefined packet-size and timing axes, which may limit its ability to capture task-specific relationships. 


A conceptually related approach was proposed by Segal et al. \cite{segal2023using}, who transformed sequences of human skeletal keypoints into a two-dimensional image representation termed TSSCI. By stacking temporally ordered skeleton descriptors and encoding spatial coordinates as color channels, the authors represented an entire movement as a single image and successfully applied CNNs for classification and similarity analysis. Although developed for human motion analysis, this work demonstrates the benefits of transforming one-dimensional sequential data into a two-dimensional representation prior to convolutional processing. Unlike TSSCI, which relies on a handcrafted transformation specific to skeletal topology, the proposed EMBD module performs a learnable inflation of traffic measurements into a two-dimensional feature space, enabling the extraction of higher-order relationships between measurements while remaining domain independent.

Similarly, in our previous work, Sidorov et al. \cite{sidorov2024revisiting}, a one-dimensional vector of tweet activity indicators was transformed into a two-dimensional matrix through a simple row-stacking operation. The resulting representation was shown to be effective for predicting future user activity and for uncovering latent connectivity patterns between users. These findings suggest that projecting sequential or tabular data into a two-dimensional space can expose structural relationships that are difficult to capture in the original one-dimensional representation, thereby improving the effectiveness of convolutional feature extraction.

The closest approach to the problem addressed in this work was proposed by Lopez-Martin et al. \cite{lopez2018deep}, who introduced a hybrid CNN–LSTM–Gaussian Process architecture for QoE prediction. The proposed method estimates QoE from passive traffic measurements, utilizing engineered statistical features derived from network packet traces, including packet size and inter-arrival time statistics, percentiles, means, and variances. Unlike conventional QoE prediction frameworks, the method does not rely on application-level information and therefore remains applicable under encrypted traffic conditions.

Nevertheless, the approach exhibits several limitations. First, the model performs binary classification of seven predefined visual artifacts, such as blur, ghosting, blockness, and chrominance distortions, rather than predicting a continuous-valued QoE metric. Consequently, the method detects the occurrence of impairments rather than directly estimating user-perceived quality. Second, the model relies on handcrafted statistical features instead of raw traffic measurements, introducing an additional feature-engineering stage that may increase processing latency and limit real-time applicability. Third, the temporal context is restricted to elementary flows spanning only three seconds, which may be insufficient to capture longer-term QoE dynamics and memory effects commonly observed in subjective quality perception. Finally, the pseudo-image representation employed by the CNN is constructed from aggregated statistical descriptors rather than the original traffic measurements. While this transformation facilitates the use of convolutional architectures, it inevitably reduces temporal resolution and may discard fine-grained dependencies present in the raw traffic stream, potentially limiting the model's ability to capture complex QoE-related patterns.

%% file: sections/methodology.tex
\section{Methodology}
\label{sec:methodology}
This section presents the proposed QoE prediction framework and provides a detailed description of its individual components. In addition, we briefly review the underlying techniques employed in each module and motivate the corresponding design choices. The section is structured into three parts: feature selection, feature preprocessing, and model architecture.

\subsection{Framework Design}

\begin{figure*}[htbp!]
    \centering
    \includegraphics[width=1.0\linewidth]{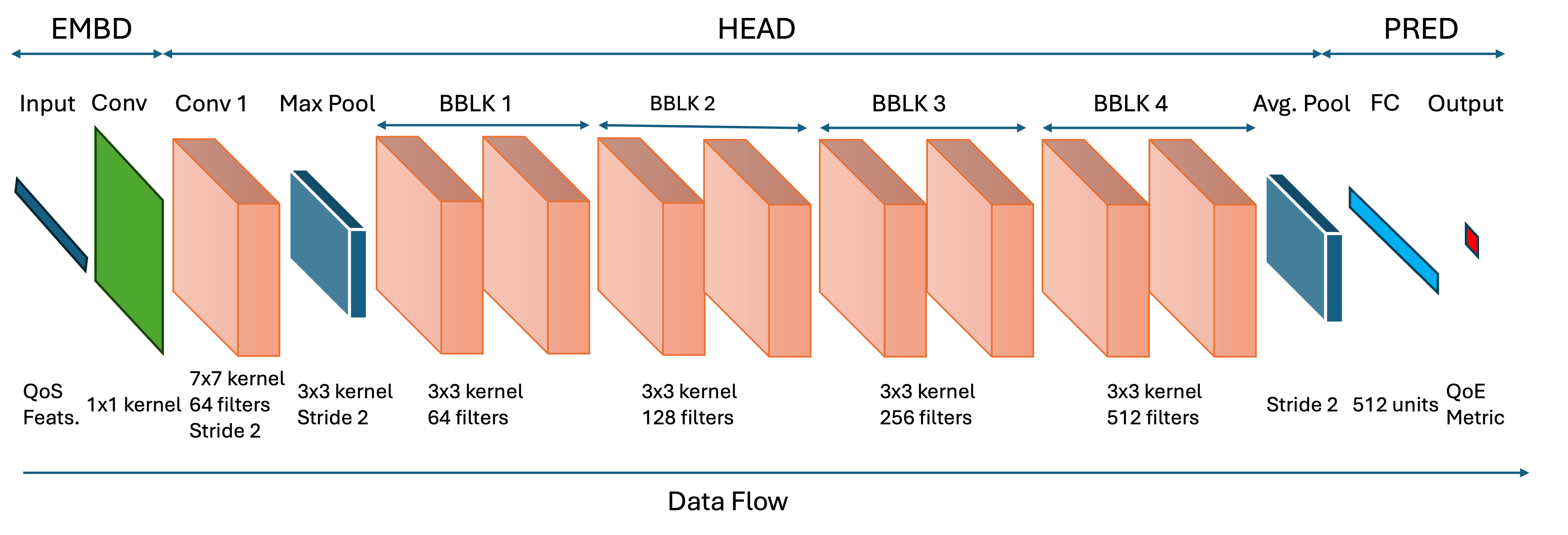}
    \caption{
    This figure illustrates the proposed qCNN architecture and its constituent modules. (a) The embedding module (EMBD), which transforms the one-dimensional QoS feature vector ($X_{1\times M}$) into an ($M \times M$) representation. (b) The head CNN module (HEAD), which receives the embedded QoS representation and performs convolution and pooling operations to extract locally informative features. (c) The predictor module (PRED), which generates the final QoE prediction based on the high-level features extracted by the HEAD module.
    }
    \label{fig:qcnn_arch}
\end{figure*}

The proposed framework consists of three modules: (1) an \textit{Embedding} (EMBD) module responsible for transforming one-dimensional QoS feature vectors extracted from the data line into two-dimensional representations; (2) a \textit{Head} (HEAD) module that extracts locally salient features from the embedded QoS representations; and (3) a \textit{Predictor} (PRED) module implemented as a fully connected layer that maps the extracted features to the target QoE metric. The overall architecture is illustrated in \cref{fig:qcnn_arch}.

The design is partially inspired by our previous work \cite{sidorov2024revisiting}, in which we introduced a composite architecture combining an autoencoder and a convolutional neural network for user activity prediction on the \textit{X} social platform. In that study, the model operated exclusively on micro-level user activity features, namely binary indicators denoting the presence (1) or absence (0) of a tweet at each timestamp, rather than on aggregated statistical descriptors. By transforming these one-dimensional activity sequences into two-dimensional representations and processing them with a CNN-based feature extractor, we demonstrated accurate prediction of future user interactions and the extraction of latent connectivity patterns between users.

The results of that study suggested that transforming one-dimensional sequential measurements into a structured two-dimensional representation can improve the ability of convolutional architectures to capture complex dependencies within the data. We hypothesize that a similar principle applies to network traffic analysis, where meaningful traffic patterns emerge from the interactions between neighboring measurements and may be associated with specific QoE outcomes.

The present work extends this idea in two important ways. First, instead of constructing two-dimensional matrices through fixed row-wise stacking, as proposed in \cite{sidorov2024revisiting}, we introduce a dedicated EMBD module that performs a learnable transformation from a one-dimensional traffic sequence into a two-dimensional feature representation. Second, we replace binary activity indicators with low-level QoS measurements extracted from encrypted traffic, namely packet sizes and packet inter-arrival times (PIATs).

The underlying assumption, supported by a substantial body of prior work \cite{chen2017seq2img, rezaei2018achieve, vu2017deep, rezaei2018deep, lotfollahi2020deep, aceto2018mobile, lopez2017network, hochst2017unsupervised}, is that communication protocols generate characteristic low-level traffic patterns that can be observed even when the payload is encrypted. These patterns often reflect protocol-specific behaviors and operational procedures, such as connection establishment, congestion control, and data transmission mechanisms. Consequently, preserving the relationships between neighboring measurements is essential for effective representation learning. By transforming traffic measurements into a two-dimensional feature space, the proposed framework enables the extraction of such patterns and their subsequent mapping to the target QoE metric.

\subsection{Embedding Module (EMBD)}
\label{subsec:embd-module}

The embedding (EMBD) (\cref{fig:qcnn_arch}(a)) module is implemented as a single convolutional (Conv.) layer with a $1 \times 1$ kernel and a number of output channels equal to the input length, $M$. Consequently, the output of this module is an $M \times M$ tensor. The module performs the inflation operation introduced in \cite{sidorov2024revisiting}. Unlike the original implementation, where inflation was achieved through simple row-wise vector stacking, the proposed approach employs a learnable convolutional transformation, thereby preserving all available training samples and avoiding the reduction in effective training data associated with the stacking procedure.

In addition to performing inflation, the EMBD module serves as an initial learnable feature transformation stage. By employing a $1 \times 1$ convolution, the input feature vector is projected into a two-dimensional $M \times M$ latent representation before being processed by the subsequent convolutional layers. This operation enables the network to learn weighted combinations and interactions among individual input features while preserving the structural information contained in the original traffic sequence.

Furthermore, unlike previous approaches that rely on fixed one-dimensional-to-two-dimensional transformations, as discussed in \cref{subsub:1d_2d}, the proposed EMBD module jointly learns the transformation with the prediction task. As a result, the generated representation is optimized with respect to the prediction objective, allowing the network to capture more complex relationships between traffic measurements than would be possible with handcrafted or static mappings.

The resulting enriched representation provides a more expressive feature space for downstream feature extraction. Consequently, subsequent convolutional layers can more effectively identify higher-level traffic patterns and dependencies, thereby enhancing the model's ability to capture traffic characteristics relevant to accurate QoE prediction.

\subsection{Head Model}
\label{subsec:head-module}
The HEAD module (\cref{fig:qcnn_arch}(b)) is responsible for extracting informative features from the embedded traffic representation. In our experiments, we employed the ResNet-18 architecture proposed by He et al. \cite{he2016deep}. ResNet-18 was selected due to its favorable trade-off between predictive performance and computational complexity, making it well suited for real-time applications where inference latency is important. Furthermore, its relatively low computational requirements make it suitable for deployment on resource-constrained edge and IoT devices, enabling QoE monitoring closer to end users rather than relying exclusively on centralized infrastructure.

In addition, the widespread adoption of ResNet-18 and the availability of pretrained ImageNet-1K weights make it particularly attractive for transfer-learning applications. Compared with deeper variants such as ResNet-34 and ResNet-50, ResNet-18 provided the best balance between computational cost and predictive performance in our preliminary experiments.

ResNet-18 is composed of a stack of convolutional residual basic blocks (BBlocks), each consisting of convolutional layers, batch normalization, and non-linear activation functions. Residual skip connections are employed to facilitate gradient propagation during training and to mitigate the degradation problem commonly observed in deeper neural networks. These connections enable the network to learn residual mappings rather than direct transformations, thereby improving optimization and feature learning.

The output of the final residual block is subsequently forwarded to the predictor module, which generates the final QoE estimate.

\subsubsection{Convolutional Neural Networks (CNNs)}
\label{subsub:cnns}
Convolutional neural networks (CNNs) are specialized neural architectures designed to extract local patterns through the application of learnable filters. During the convolution operation, small two-dimensional kernels are repeatedly applied across the input representation, producing feature maps that capture relevant local structures. These operations are typically followed by pooling layers, such as max pooling, which reduce the dimensionality of the representation while retaining the most salient features.

One of the key properties of CNNs is their ability to detect similar patterns regardless of their exact spatial location within the input. This characteristic is particularly advantageous in the proposed framework, as traffic-related temporal patterns may occur at different positions within the embedded traffic representation. By exploiting local connectivity and weight sharing, the HEAD module is able to identify recurring traffic structures and transform them into high-level features that are subsequently used for QoE prediction.

\subsection{Predictor}
\label{subsec}
The PRED module (\cref{fig:qcnn_arch}(c)) is responsible for generating the final QoE estimate. It consists of a fully connected (FC) layer with 512 hidden units followed by a single-neuron output layer. The module receives the flattened feature maps produced by the HEAD module and performs the final feature-to-target mapping, generating a scalar output corresponding to the predicted QoE metric.

\subsection{Data Preparation for the Training and Testing Pipeline}

\begin{figure*}[htbp!]
\centering
\includegraphics[width=1.0\linewidth]{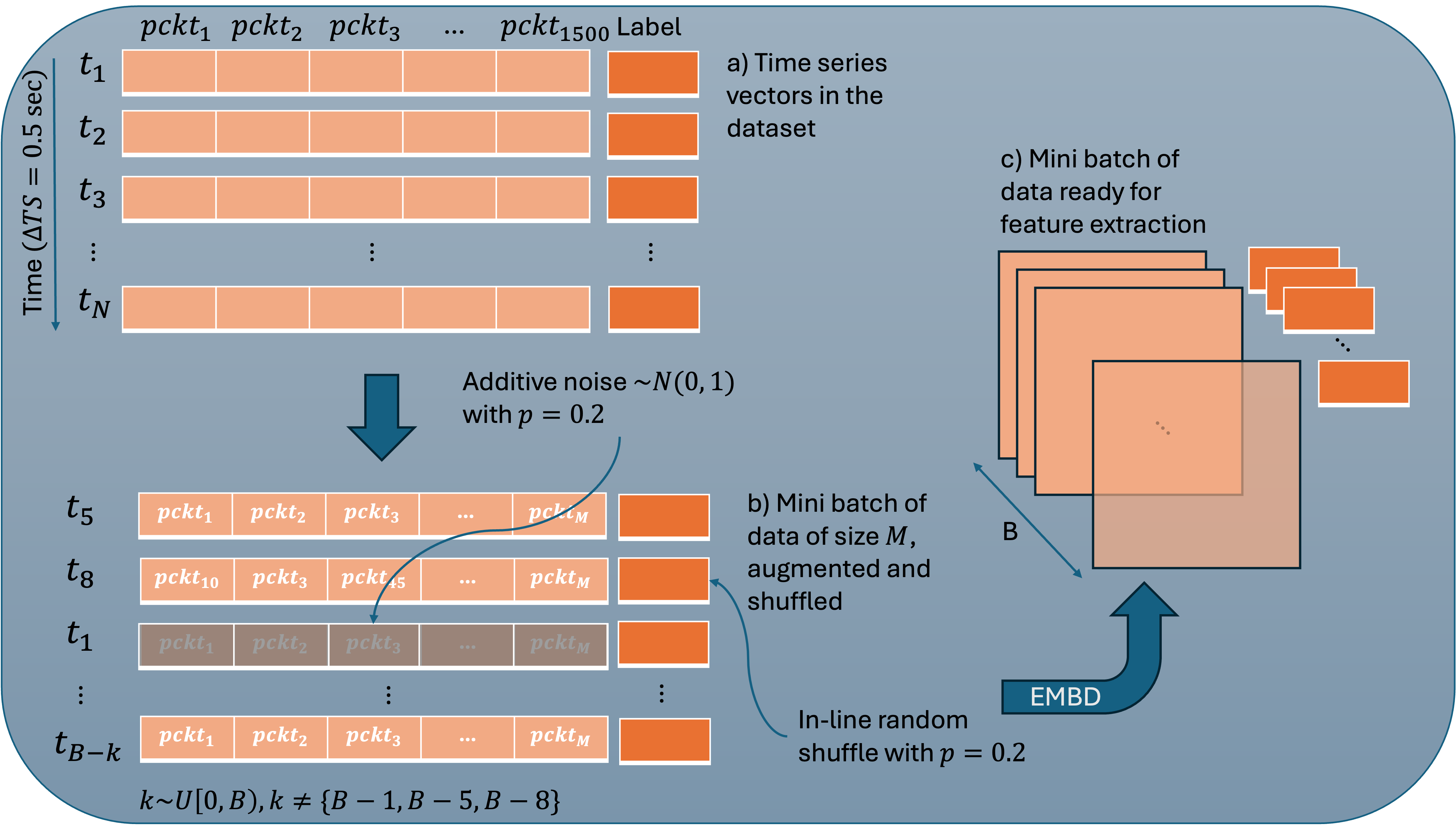}
\caption{Illustration of the proposed data processing pipeline, from raw traffic measurements to the embedded two-dimensional representation generated by the EMBD module and subsequently processed by the CNN-based feature extractor.}
\label{fig:data-pipe}
\end{figure*}

Since the features contained in the dataset correspond to QoS measurements collected at consecutive timestamps, the resulting data constitute a time series. To enable processing by the CNN-based feature extractor, the one-dimensional sequence must be transformed into a two-dimensional representation, as described in \cref{subsec:embd-module}.

During both training and inference, the data are processed in batches. For each iteration, a batch ($B$) consisting of samples of length ($M$) is extracted from the dataset. During training, data augmentation is applied with probability $p=0.2$ using two independent operations: (i) additive Gaussian noise, where each sample is perturbed by noise drawn from $N(0,1)$, and (ii) random permutation of elements within the sample. These augmentation techniques are intended to improve the robustness and generalization capability of the model.

Following augmentation, the one-dimensional traffic sequences are transformed into two-dimensional representations through the EMBD module using the inflation procedure introduced in \cite{sidorov2024revisiting}. The resulting $M \times M$ representations are subsequently passed to the HEAD module for feature extraction. The overall data processing workflow is illustrated in \cref{fig:data-pipe}.

\subsection{Stationarity}
\label{sub:stationarity}
An auto-regressive (AR) process is a stochastic process whose output variable depends linearly on its own previous values, and some stochastic noise process, and may be described by the following relation

\begin{equation}
    X_t = \sum_{\tau=1}^T \varphi_\tau X_{t-\tau} + \varepsilon_t
    \label{eq:ar-proccess}
\end{equation}

\noindent where $\varphi_\tau$ represents the parameters of the model at time $\tau$, $X_{t-\tau}$ represents the output of the model in time $t-\tau$, and $\varepsilon_t \sim N(0, \sigma^2)$, a white noise process with zero mean and constant variance $\sigma^2$.

Stationarity refers to the degree to which a signal exhibits consistent statistical properties over time. Formally, a signal $x_t$ is considered stationary if its distribution at time $t$ is the same as its distribution at time $t+\Delta t$, for any arbitrary offset $\Delta t$. This property is particularly important in our context, as stationary signals tend to contain recurring patterns that may hold strong predictive relevance.

The Dickey-Fuller (DF) \cite{dickey1979distribution} test is used as a test of stationarity for Markovian AR processes, where each future output at time $t$ depends only on its immediate previous output at time $t-1$, and is described as 

\begin{equation}
    X_t = \rho X_{t-1} + \varepsilon_t
    \label{eq:markov-process}
\end{equation}

\noindent where $X_t$ is the variable of interest in time $t$, $\rho$ is the coefficient, and $X_{t-1}$ is the value of the variable of interest in the previous timestamp. In this simplified version of an AR process, we may test for the presence of a unit root by transforming \cref{eq:markov-process} to the following form:

\begin{equation}
    X_t - X_{t-1} = \rho X_{t-1} + \varepsilon_t - X_{t-1}
    \label{eq:unit-root-mid}
\end{equation}

\begin{equation}
    \Delta X_t = (\rho - 1)X_{t-1} + \varepsilon_t = \delta X_{t-1} + \varepsilon_t
    \label{eq:unit-root}
\end{equation}

\noindent where $\Delta$ represents the first difference operator, $\delta \equiv \rho - 1$. In case $\rho = 1$ is a root of \cref{eq:unit-root}, the model becomes dependent solely on the behavior of the $\varepsilon_t \sim N(0, \sigma^2)$ term, and hence non-stationary.

To test for the stationarity, the DF test poses two hypotheses, viz. the null hypothesis ($H_0$) --- a unit root is present in the sample, and the alternative hypothesis ($H_1$) --- a unit root does not appear among the solutions of the characteristic equation of the process, and the signal is stationary. The test rejects the null hypothesis if the signal statistic is smaller than the statistic corresponding to one of the significance levels, viz. 1\%, 5\%, and 10\% presented in \cref{tab:df-table} \cite{dickey1979distribution}.

The intuition behind the regular DF test is that if the process is stationary it should return to the mean value in the course of its advancement, i.e., large values should be followed by small and vice-a-versa. The constant return to the mean will make the current level of the series a significant predictor of the value in the next timestamp, and the null hypothesis will be rejected. On the other hand, if small and large values occur with the same probability it means that the series is non-stationary, and we may not be able to dismiss the null hypothesis, concluding non-stationarity of the AR process.

\begin{table}[htbp!]
    \begin{center}
    \caption{Dickey--Fuller critical values \cite{dickey1979distribution} used in the stationarity analysis. The null hypothesis ($H_0$: the series contains a unit root) is rejected when the test statistic $\tau$ is smaller than the critical value corresponding to the chosen significance level.}
    \label{tab:df-table}
            \begin{tabular}{| c | c | c | c |}
                \hline
                    \diagbox[width=10em]{N}{Confidence}    & 1\%  & 5\% & 10\% \\
                \hline
                    25            & -3.724        & -2.986       & -2.633\\ 
                \hline
                    100           & -3.498        & -2.891       & -2.582\\ 
                \hline
                    500+          & -3.434        & -2.863       & -2.568\\
                \hline
            \end{tabular}
    \end{center}
\end{table}

As the dependence of the next packet size only on the packet size of the last packet to arrived may not be asserted with a sufficient confidence, in our experiments we used the Augmented Dickey-Fuller (ADF) test, which differs from the DF test by addition of lag variables to the characteristic equation of the AR process as follows

\begin{equation}
    \Delta X_t = \rho X_{t-1} + \beta t + \alpha + \sum_{i=1}^{L}\delta y_{t-i} + \varepsilon_t
    \label{eq:adf_characteristic_equation}
\end{equation}

\noindent where $\alpha$ is some constant, $\beta$ is the coefficient on a time trend, and $L$ represents the lag order of the AR process (i.e., how big is the history that the next value of the process depends on). This formulation of the AR process allows consideration of higher order of AR models, and not just AR(1), as in the case of the regular DF test.

Intuitively, the ADF test states that if a process is characterized by a unit-root (i.e., nonstationary), then $X_{t-1}$ should provide no new information regarding the changes in the next timestamp, thus we would conclude that $\rho = 0$, and won't be able to reject $H_0$. On the other hand if the process has no unit root in the set of the solutions for its' characteristic equation, the process will revers to the mean, i.e., the lagged level will provide relevant information for predicting the next change in the value, which will reject the $H_0$.

%% file: sections/datasets.tex
\section{Data}
\label{sec:dataset}
In this section, we describe the datasets used in this study, including the data collection procedure, feature extraction process, labeling methodology, and the motivations that guided the design of the experimental setup.

\subsection{Data Collection}
\label{subsec:data-collect}
\begin{figure}[htbp!]
\centering
\includegraphics[width=1.0\linewidth]{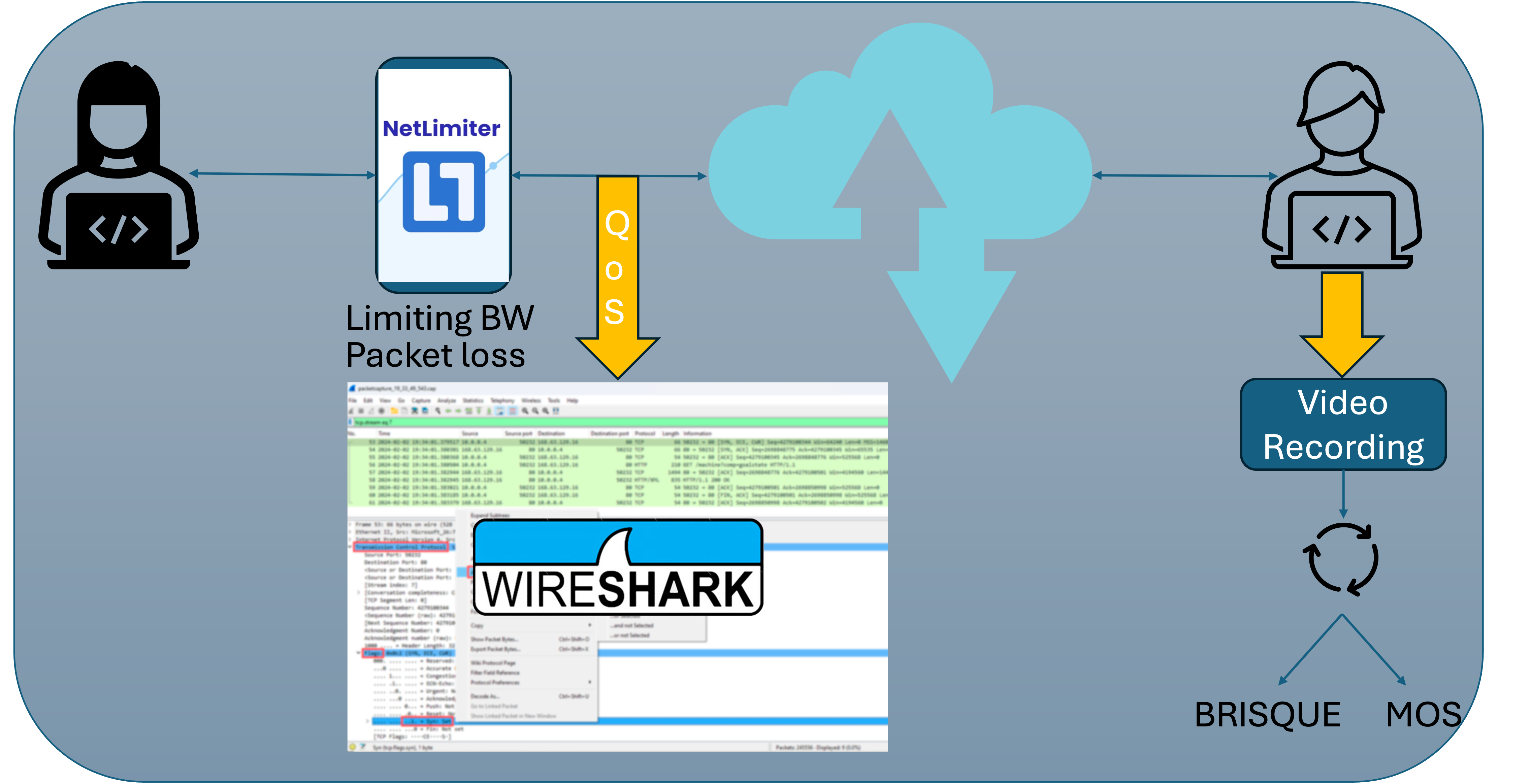}
\caption{Schematic representation of the data collection setup. Two users participated in a video conferencing (VC) session. On the laboratory side, the available bandwidth was controlled using the NetLimiter and Clumsy applications to emulate network impairments. Traffic measurements were collected using Wireshark, while the video stream was recorded on the remote client for subsequent QoE label extraction using BRISQUE and MOS metrics.}
\label{fig:data-collect-setup}
\end{figure}

In the present study, we utilized two custom datasets collected under controlled experimental conditions on the WhatsApp (WA) and Zoom (ZO) platforms, following the general principles described in \cite{sidorov2025reliable}. During data collection, video conferencing (VC) calls were established between two users communicating over the Internet: one using a stationary computer located in the laboratory and the other connecting from a geographically remote location. This configuration ensured that the traffic traversed external networks, thereby emulating realistic operating conditions.

During each call, Quality of Service (QoS) measurements were collected on the laboratory computer, while video recording and QoE label extraction were performed on the remote client. Packet sizes (PCKT) were monitored for both incoming and outgoing traffic. Since each packet was associated with a timestamp, packet inter-arrival times (PIATs) were also extracted.

Given that the maximum transmission unit (MTU) in Ethernet networks is 1500 bytes, traffic traces were represented using sequences of 1500 consecutive packets. Furthermore, as Wireshark records timestamps with millisecond precision, the extracted measurements were aggregated over 0.5-second intervals to reduce measurement noise and improve temporal stability. A schematic representation of the data collection process is shown in \cref{fig:data-collect-setup}. 

\subsection{WhatsApp Dataset (WA)}

The WhatsApp (WA) dataset comprises 107 video conferencing (VC) sessions, each lasting four minutes. QoS features were extracted from the captured traffic according to the procedure described in \cref{subsec:data-collect}. The recorded videos were labeled using the Blind/Referenceless Image Spatial Quality Evaluator (BRISQUE) \cite{mittal2011blind}, an objective no-reference image quality metric based on deviations from natural scene statistics. BRISQUE scores range from 0 to 100, with lower values corresponding to higher visual quality.

Network impairments were introduced using the NetLimiter and Clumsy applications to control the available bandwidth ($B$) and inject packet loss. Bandwidth limitations were applied using two configurations. In the first configuration, a constant bandwidth was randomly selected from the set $B \in \{15, 30, 60, 125, 250\}$ kBps. In the second configuration, the bandwidth was initially set to
$B_{\mathrm{init}} = 250$ kBps, and subsequently reduced to a randomly selected lower value $B_{\mathrm{low}} \sim U[10,150]$ kBps.

This configuration was designed to emulate realistic network degradation scenarios, in which the available bandwidth decreases dynamically rather than remaining constant throughout the call. Additionally, packet loss was independently introduced with probabilities $p_{\mathrm{loss}} \in \{0.00, 0.01, 0.02, 0.03, 0.10\}$.

Following feature extraction and labeling, the final dataset contained 10,607 samples.

\subsection{Zoom Dataset (ZO)}

The Zoom (ZO) dataset was considerably smaller and consisted of 30 video clips generated from five 6-second video conferencing (VC) sessions conducted under varying bandwidth conditions, $B \in \{20, 40, 60, 80, 100, 250\}$ kBps.

The videos were labeled using the Mean Opinion Score (MOS) QoE metric in accordance with the recommendations provided in \cite{recommendation20081080}. Since MOS directly reflects users' subjective perception of service quality and is widely regarded as the gold standard for QoE assessment, this dataset was collected to evaluate the practical applicability of the proposed approach and its ability to generalize beyond objective quality metrics to human-centered QoE measurements.

\subsection{Features}
\begin{figure*}[htbp!]
    \centering
    \begin{subfigure}[t]{0.45\linewidth}
        \centering
        \includegraphics[width=1.0\linewidth]{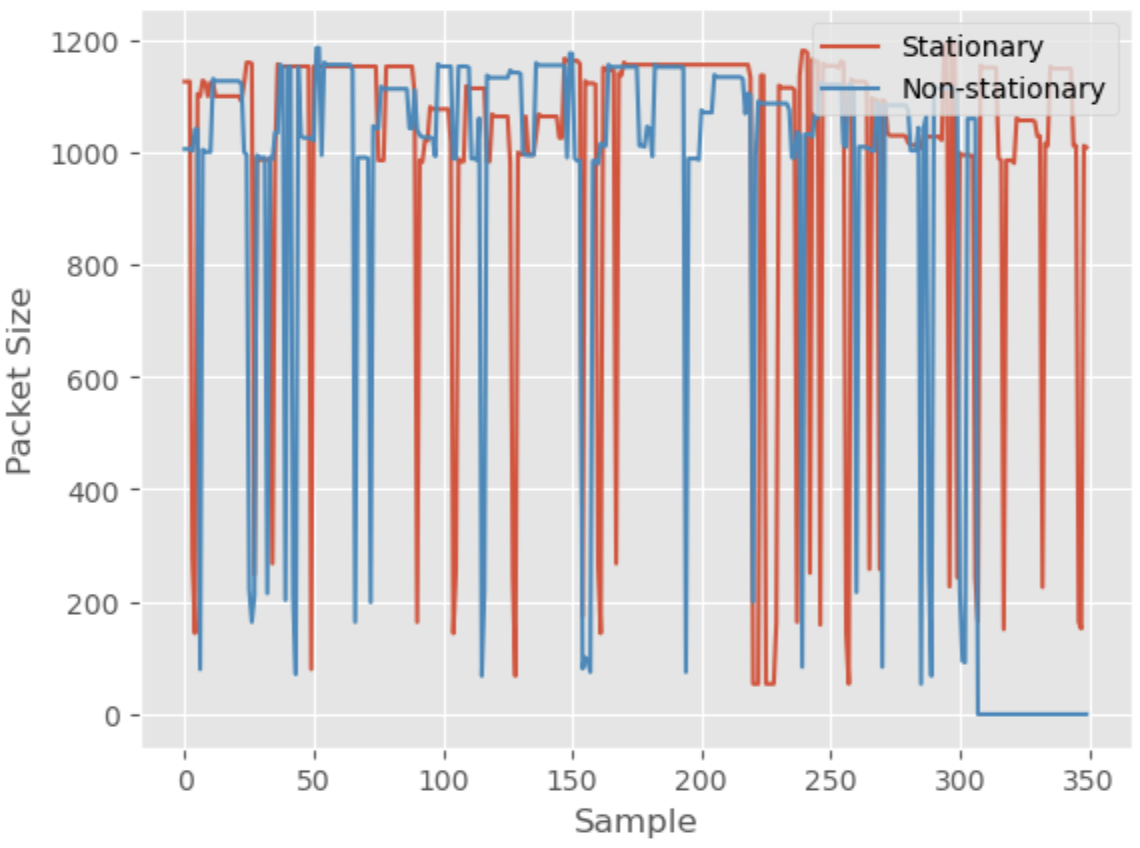}
        \caption{\label{subfig:wa-pckt-sig-stat-nonstat}Examples of stationary (red) and non-stationary (blue) PCKT signals from the WA dataset. }
    \end{subfigure}
    \begin{subfigure}[t]{0.45\linewidth}
        \centering
        \includegraphics[width=1.0\linewidth]{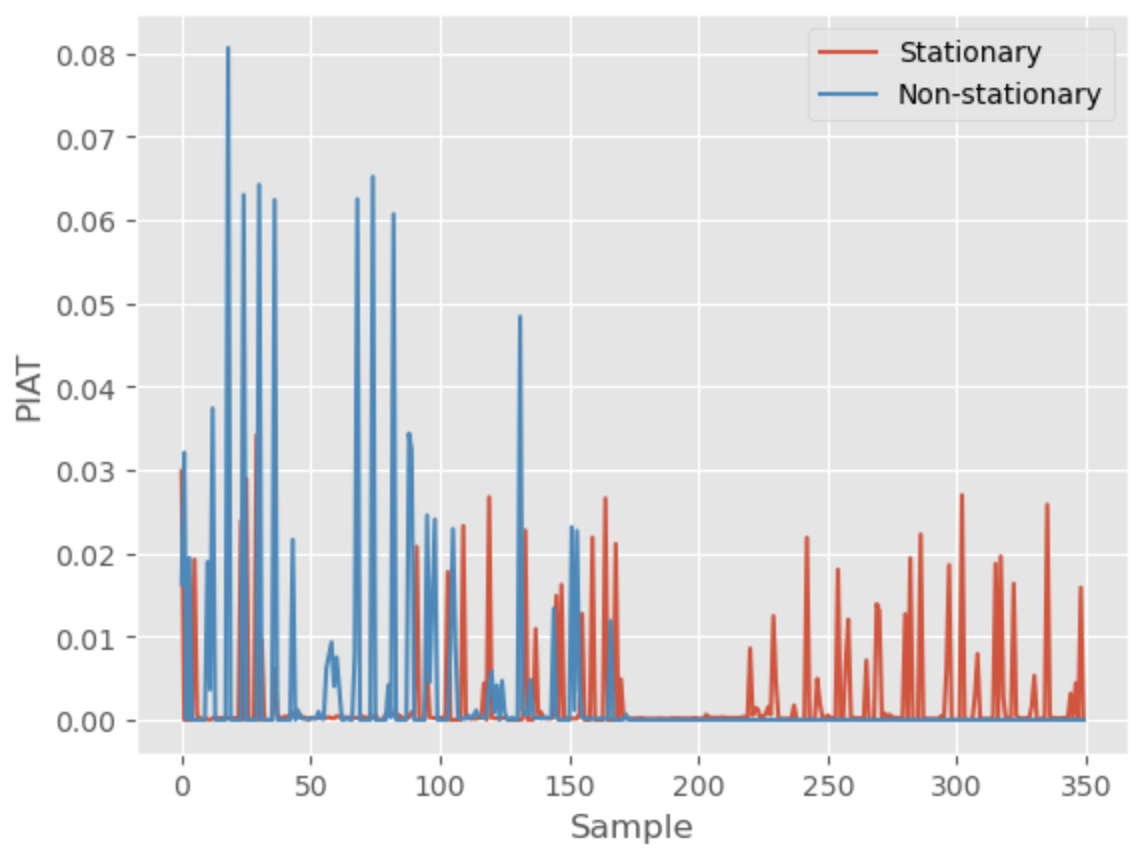}
        \caption{\label{subfig:zo-piat-sig-stat-nonstat}Examples of stationary (red) and non-stationary (blue) PIAT signals from the WA dataset. }
    \end{subfigure}
    \begin{subfigure}[t]{0.45\linewidth}
        \centering
        \includegraphics[width=1.0\linewidth]{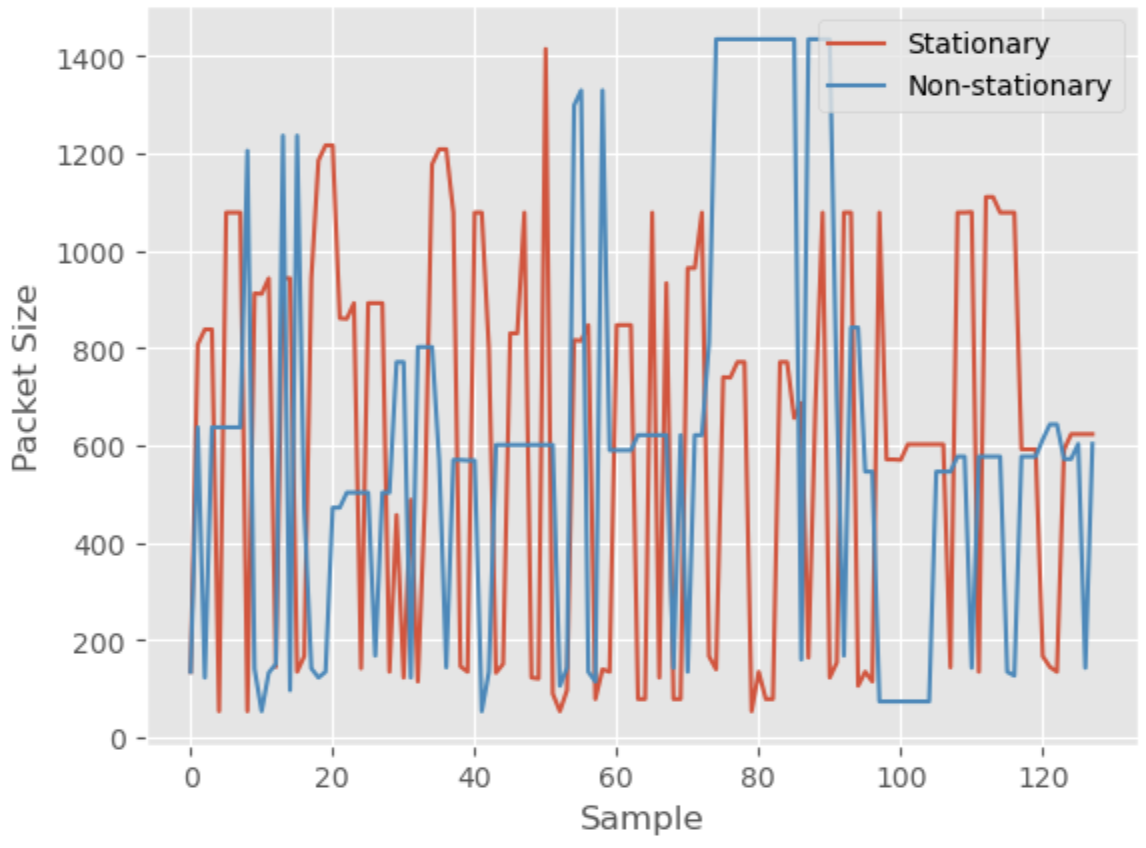}
        \caption{\label{subfig:zo-pckt-sig-stat-nonstat}Examples of stationary (red) and non-stationary (blue) PCKT signals from the ZO dataset. }
    \end{subfigure}
    \begin{subfigure}[t]{0.45\linewidth}
        \centering
        \includegraphics[width=1.0\linewidth]{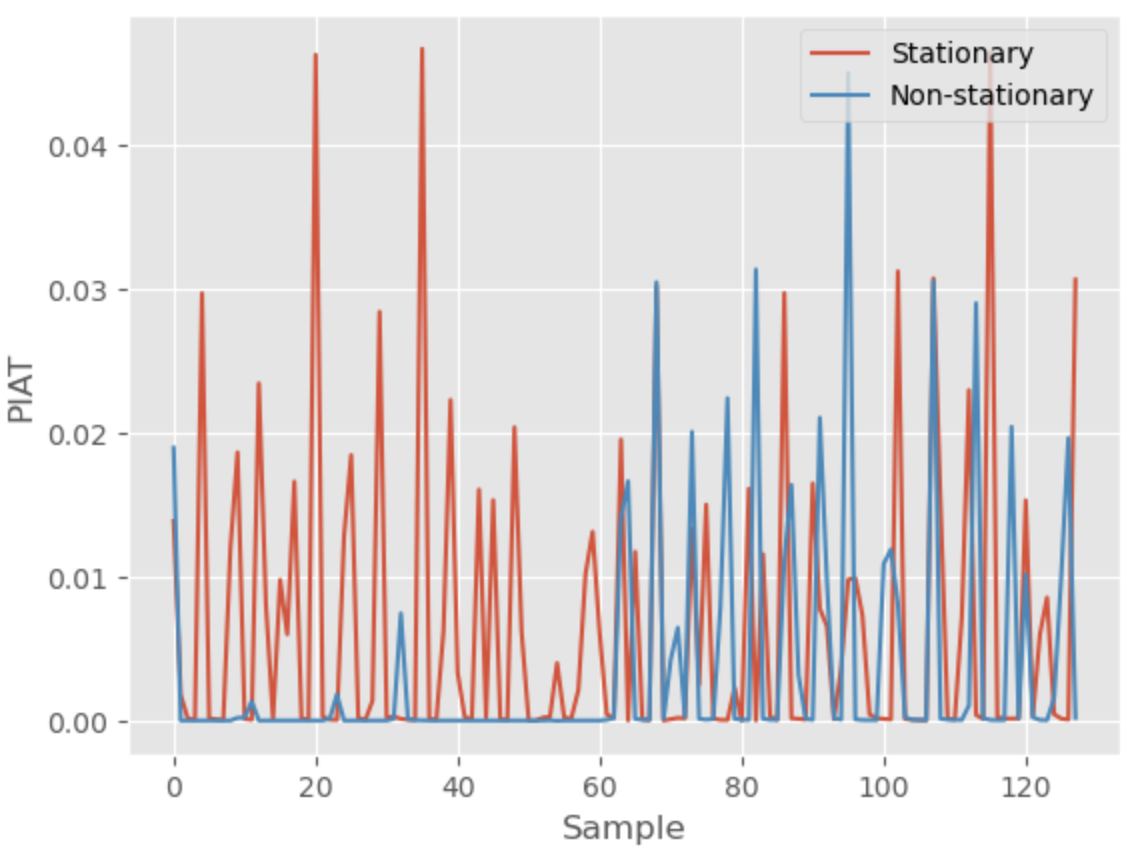}
        \caption{\label{subfig:zo-piat-sig-stat-nonstat}Examples of stationary (red) and non-stationary (blue) PIAT signals from the ZO dataset. }
    \end{subfigure}
    \caption{\label{fig:signal-examp}Examples of PCKT and PIAT signals extracted from the WA (a, b) and ZO (c, d) datasets. In each case, the red signal corresponds to a stationary sample, whereas the blue signal corresponds to a non-stationary sample. The prolonged sequence of trailing zero-valued observations in the blue signals contributes to the rejection of the stationarity assumption.}
\end{figure*}

Motivated by the successful use of micro-scale features in \cite{sidorov2024revisiting}, and by the fact that such features correspond to direct measurements of the communication channel and can therefore be extracted with minimal processing overhead, we focus on low-level traffic characteristics as the primary source of information for QoE prediction. This choice is additionally supported by a substantial body of prior work demonstrating that packet size is one of the most informative features for encrypted traffic analysis and classification \cite{orsolic2017machine, panchenko2016website, dimopoulos2016measuring, bronzino2019inferring, gutterman2019requet, oura2024qoe, shen2020deepqoe, wassermann2019see, lopez2018deep}.

Accordingly, in the present study we use packet size (PCKT) and packet inter-arrival time (PIAT) measurements aggregated over 0.5-second intervals as the primary input features. This process resulted in a total of 51,341 labeled samples for the WA dataset and 500 labeled samples for the ZO dataset. Examples of the resulting feature signals are shown in \cref{fig:signal-examp}.

We hypothesize that PCKT and PIAT provide meaningful information regarding the state of a video conferencing (VC) session, including whether the communication is stable, degraded, temporarily frozen, or undergoing reconnection. Under favorable network conditions, packets tend to flow with minimal disruption and larger packet sizes are typically observed, as the video-conferencing application can reduce compression and improve visual quality. Conversely, under constrained bandwidth conditions, stronger compression is applied, resulting in lower-quality video frames and reduced packet sizes. Similarly, network impairments may influence packet timing characteristics, making PIAT measurements informative indicators of communication quality.

These observations suggest a relationship between low-level traffic measurements and the operational state of the communication channel. Consequently, PCKT and PIAT features may serve as indirect indicators of perceptual quality and can potentially be leveraged for QoE estimation without requiring access to application-layer information.



\subsection{Labels}
\label{subsec:labels}
In the present study, we focused on two QoE metrics, namely Mean Opinion Score (MOS) \cite{recommendation20081080} and Blind/Referenceless Image Spatial Quality Evaluator (BRISQUE) \cite{mittal2011blind}. MOS is widely regarded as the most reliable QoE indicator, as it directly reflects user-perceived video quality and overall satisfaction with the service.

Furthermore, as demonstrated in our previous work \cite{sidorov2025estimating}, video conferencing systems often degrade spatial quality (image quality) before reducing temporal quality (frame rate), since maintaining conversational continuity is generally prioritized by adaptive streaming algorithms. Consequently, degradation in image quality may serve as an early indicator of future QoE deterioration, enabling Internet Service Providers (ISPs) or network operators to take corrective actions before the user experience is significantly affected.

Motivated by this observation, we chose to focus on image quality degradation rather than frame rate degradation. To this end, we employed BRISQUE as the target quality metric for the WA dataset. BRISQUE is a no-reference image quality assessment metric (NR-IQAM) that estimates perceptual image quality directly from image content without requiring a pristine reference image. Compared with frame rate extraction, which often relies on specialized tools or platform-specific APIs that may not be accessible to network operators, BRISQUE can be efficiently computed from captured video frames and has been shown to exhibit strong correlation with subjective quality measures such as MOS. In addition, its relatively low computational complexity makes it well-suited for real-time quality monitoring and assessment.

\subsubsection{Blind/Referenceless Image Spatial Quality Evaluator (BRISQUE)}
\label{subsub}

BRISQUE is a no-reference image quality assessment metric that relies on natural scene statistics (NSS) extracted directly from the spatial domain of an image. The metric is based on the observation that undistorted natural images exhibit regular statistical properties, whereas distortions introduce measurable deviations from these statistics.

First, the image is normalized using mean-subtracted contrast normalization (MSCN) coefficients defined as

\begin{equation}
\hat{I}(i, j) = \frac{I(i, j) - \mu(i, j)}{\sigma(i, j) + 1},
\label{eq}
\end{equation}

\noindent where $I(i,j)$ denotes the intensity of the pixel at spatial location $(i,j)$, for $i \in {1,2,\ldots,M}$ and $j \in {1,2,\ldots,N}$ in an image of dimensions $M \times N$. The local mean intensity is computed as

\begin{equation}
\mu(i, j) = \sum_{k=-K}^{K}\sum_{l=-L}^{L} w_{k,l} I(i+k, j+l),
\label{eq}
\end{equation}

\noindent and the local contrast is estimated as

\begin{equation}
\sigma(i, j) =
\sqrt{
\sum_{k=-K}^{K}\sum_{l=-L}^{L}
w_{k,l}
\left[I(i+k, j+l)-\mu(i,j)\right]^2
},
\label{eq}
\end{equation}

\noindent where $\mu(i,j)$ and $\sigma(i,j)$ represent the local mean intensity and local contrast, respectively. The weighting coefficients $w={w_{k,l}}$ correspond to a two-dimensional circularly symmetric Gaussian kernel sampled over three standard deviations ($K=L=3$) and normalized to unit volume.

The resulting MSCN coefficients and their pairwise products are used to extract NSS features that characterize the statistical properties of the image. These features are subsequently compared against statistical models learned from large collections of pristine and distorted natural images. Distortions such as blur, JPEG compression artifacts, ringing, and transmission-related impairments alter the NSS characteristics, allowing BRISQUE to quantify the degree of image degradation.

Finally, the extracted NSS features are supplied to a support vector machine (SVM) regressor trained on human quality judgments. The output of the regressor constitutes the final BRISQUE score, where lower values correspond to higher perceptual image quality and higher values indicate more severe distortions.

\subsubsection{Mean Opinion Score (MOS)}
\label{subsub:mos}
Due to its subjective nature, MOS does not possess an intrinsic mathematical formulation. Instead, it is derived from the recommendations provided in \cite{recommendation20081080}, which define MOS as the arithmetic mean of quality ratings assigned by human observers during controlled subjective evaluation experiments. In such experiments, a video sequence is presented to a group of participants, who are asked to assess its perceived quality using a five-point scale. Specifically, the rating assigned by user $i$ is denoted by $r_i \in \{1:\text{ Bad}, 2:\text{ Poor}, 3:\text{ Fair}, 4:\text{ Good}, 5:\text{ Excellent}\}$.

The corresponding MOS value for a video $v$ is then computed as

\begin{equation}
MOS_v = \frac{1}{N}\sum_{i=1}^{N} r_i,
\label{eq:mos-comp}
\end{equation}

\noindent where $N$ denotes the number of participants who evaluated video $v$. The resulting score provides a quantitative estimate of the overall perceived quality of the video and is widely regarded as one of the most reliable measures of Quality of Experience (QoE).

Although MOS can be computed as the arithmetic mean of ratings, as defined in \cref{eq}, obtained from any number of participants, ITU-T recommendations for subjective quality assessment suggest that each test condition should be evaluated by at least 24 subjects in controlled laboratory environments and by at least 35 subjects in uncontrolled environments to ensure statistical reliability.

\subsection{Exploratory Data Analysis (EDA)}
\label{subsec:data-eda}
In this subsection, we analyze the characteristics of the WA and ZO datasets used in this study and discuss the implications of these findings for feature selection, model design, and training strategy.

\begin{figure*}[htbp!]
    \centering
    \begin{subfigure}[t]{0.32\linewidth}
        \centering
        \includegraphics[width=1.0\linewidth]{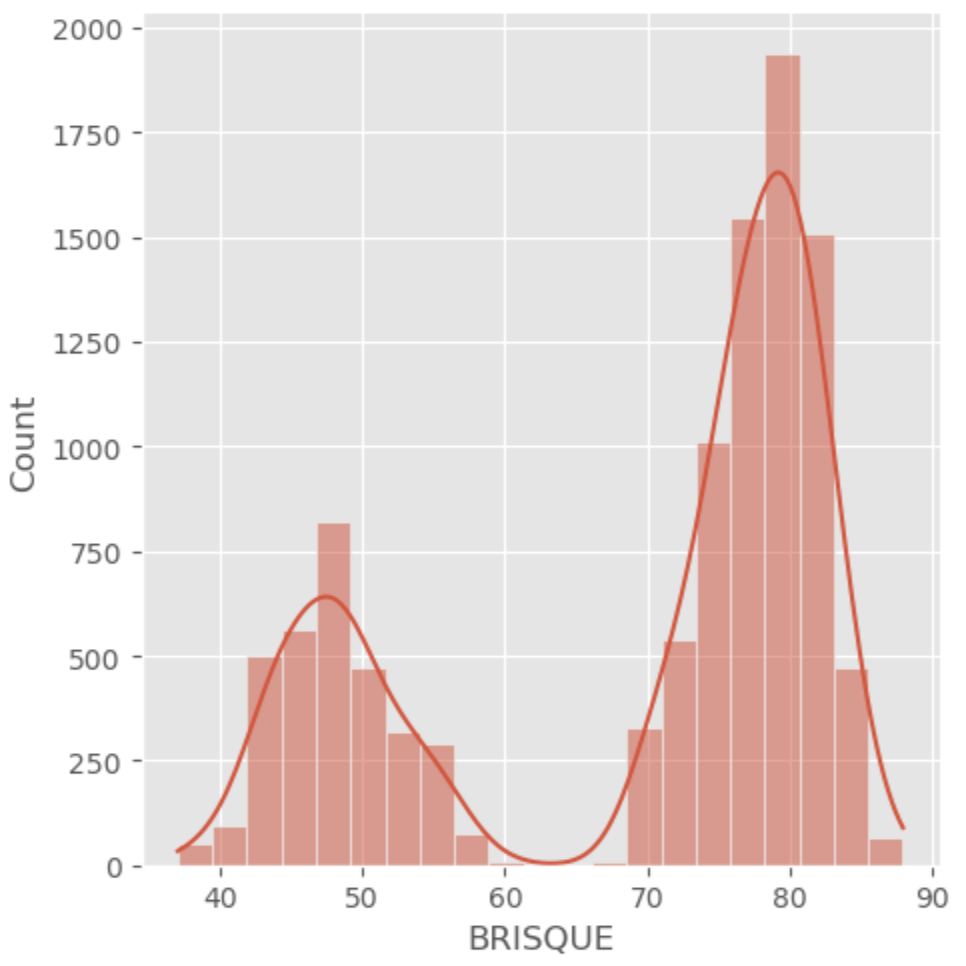}
        \caption{\label{subfig:brisque-dist}Distribution of the BRISQUE labels in the WA dataset. The distribution is markedly non-uniform and exhibits a gap in the mid-range values..}
    \end{subfigure}
    \begin{subfigure}[t]{0.32\linewidth}
        \centering 
        \includegraphics[width=1.0\linewidth]{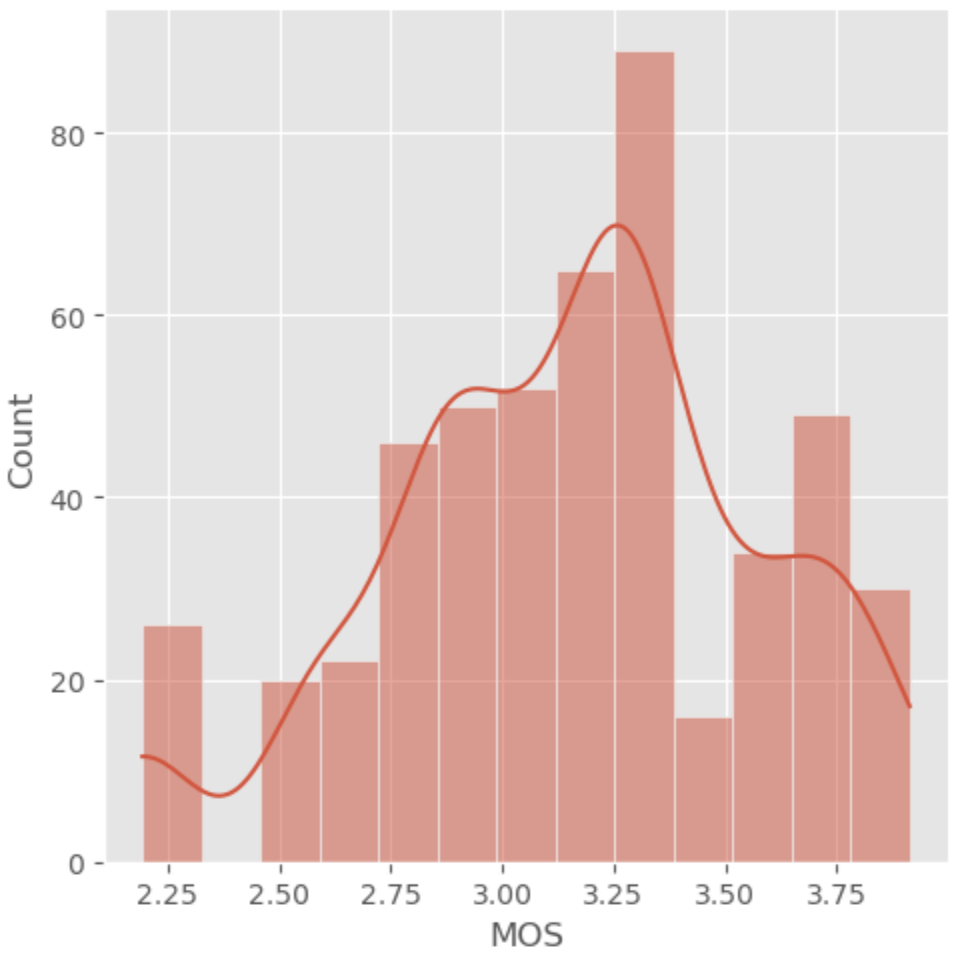}
        \caption{\label{subfig:mos-dist}Distribution of the MOS labels in the ZO dataset.}
    \end{subfigure}
    \caption{\label{fig:label-dist}Distribution of the target QoE labels used in this study, namely BRISQUE for the WA dataset and MOS for the ZO dataset. Note the pronounced non-uniformity of both distributions, which reflects the inherent difficulty of generating video conferencing sessions with predefined QoE levels under controlled experimental conditions.}
\end{figure*}

\subsubsection{Label distributions}
\label{subsub:label-dist}
The distribution of BRISQUE labels in the WA dataset exhibits a bimodal structure, with a dominant peak around a score of 50 and a secondary peak near 80 (\cref{subfig:brisque-dist}). Since lower BRISQUE values correspond to higher visual quality, the distribution suggests that the majority of the collected VC sessions were characterized by relatively good visual quality, while fewer samples corresponded to severely degraded conditions. This observation implies that prediction performance may be less reliable for lower-quality sessions due to the smaller number of available training samples, which is consistent with observations reported in \cite{lopez2018deep}. In particular, the interval between BRISQUE values of approximately 60 and 70 is sparsely populated, indicating a lack of representative training examples in this region (\cref{subfig:mos-dist}).

The distribution of MOS labels in the ZO dataset exhibits substantially higher variability (\cref{subfig:mos-dist}). This behavior is expected because MOS is a subjective quality measure that reflects individual user perception and is therefore inherently more variable than objective image-quality metrics such as BRISQUE. Furthermore, the relatively small size of the ZO dataset contributes to irregularities in the observed distribution. Consequently, the MOS labels do not appear to follow a simple parametric distribution, such as a Gaussian distribution or a mixture of Gaussians, unlike the BRISQUE labels observed in the WA dataset.

\subsubsection{Number of features}
\label{subsub:features}

\begin{figure*}[htbp!]
    \centering
    \begin{subfigure}[t]{0.45\linewidth}
        \centering
        \includegraphics[width=0.8\linewidth]{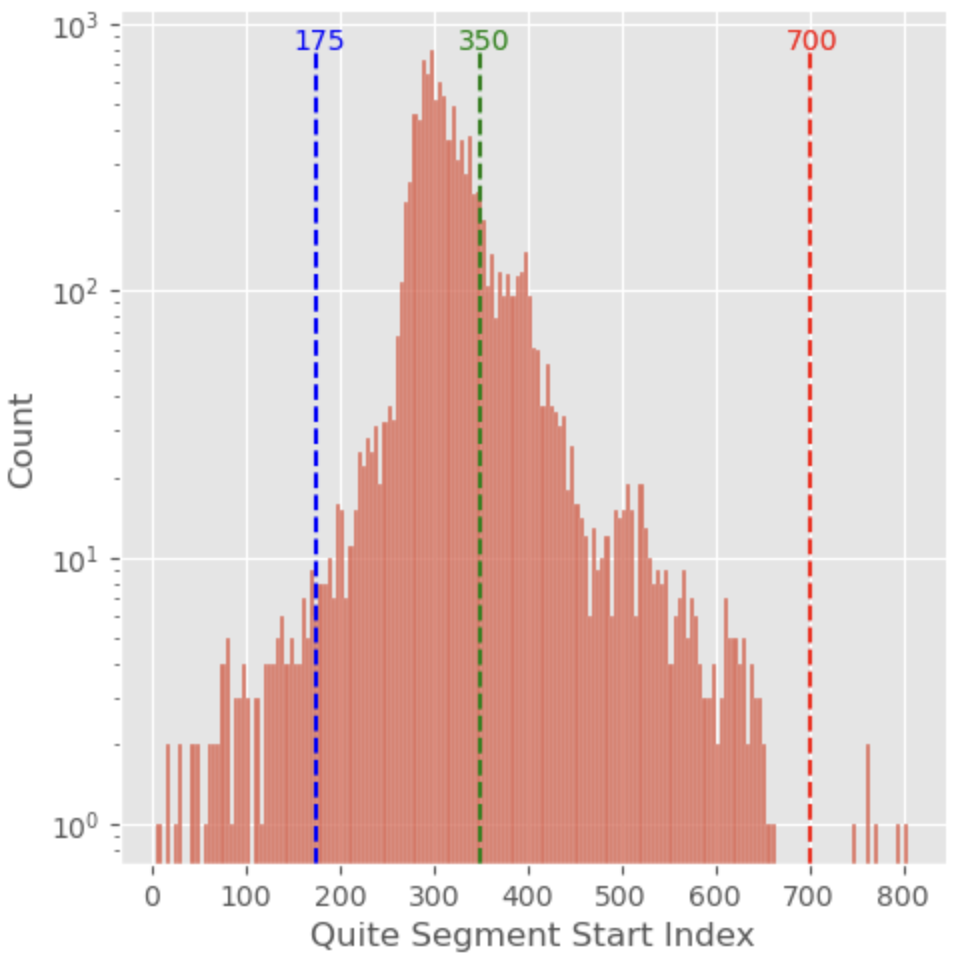}
        \caption{\label{subfig:wa-zrs-hist}Histogram of the starting positions of quiet segments identified in the PCAP recordings of the WA dataset.}
    \end{subfigure}
    \begin{subfigure}[t]{0.45\linewidth}
        \centering
        \includegraphics[width=1.0\linewidth]{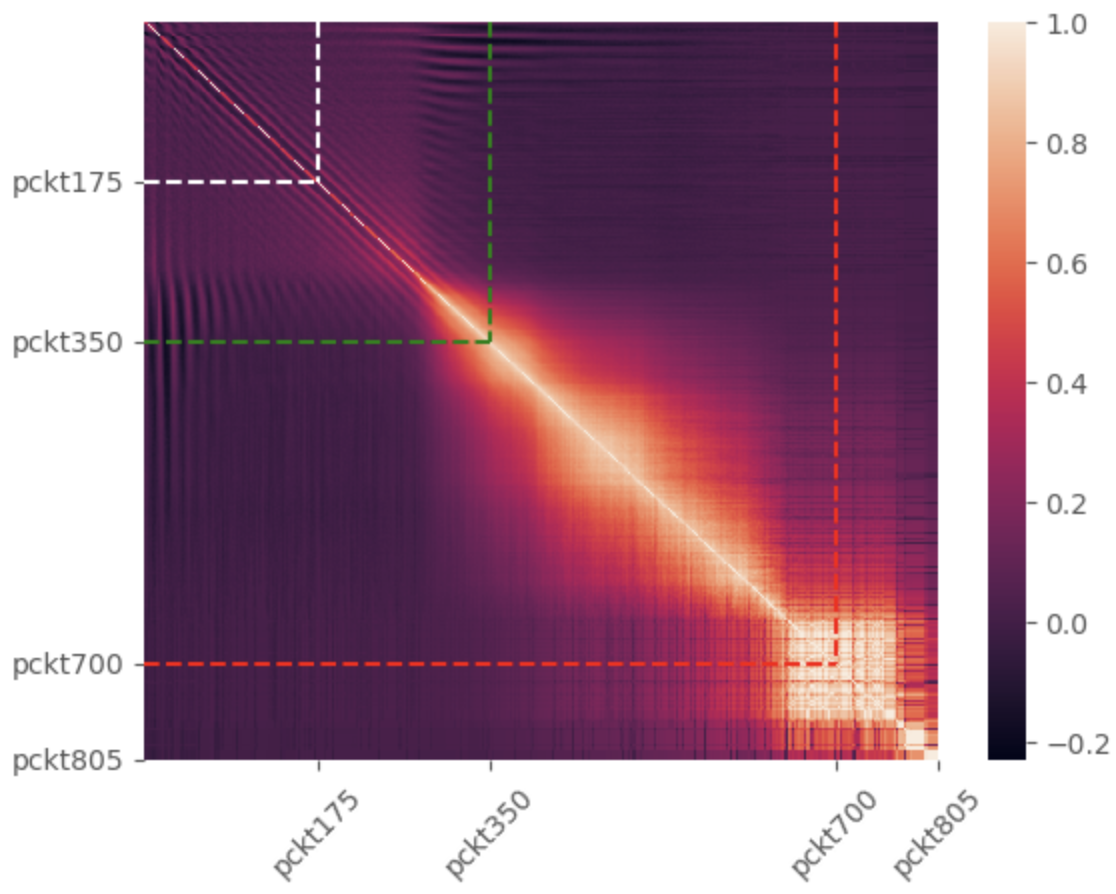}
        \caption{\label{subfig:wa-feat-corr}Correlation heatmap of the features extracted from the WA dataset.}
    \end{subfigure}
    \begin{subfigure}[t]{0.45\linewidth}
        \centering 
        \includegraphics[width=0.9\linewidth]{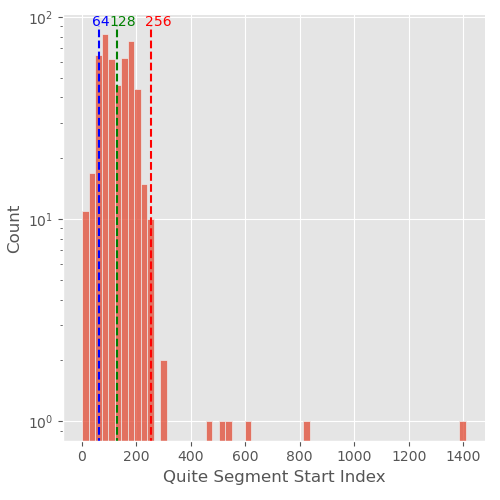}
        \caption{\label{subfig:zo-zrs-hist}Histogram of the starting positions of quiet segments identified in the PCAP recordings of the ZO dataset.}
    \end{subfigure}
    \begin{subfigure}[t]{0.45\linewidth}
        \centering 
        \includegraphics[width=1.05\linewidth]{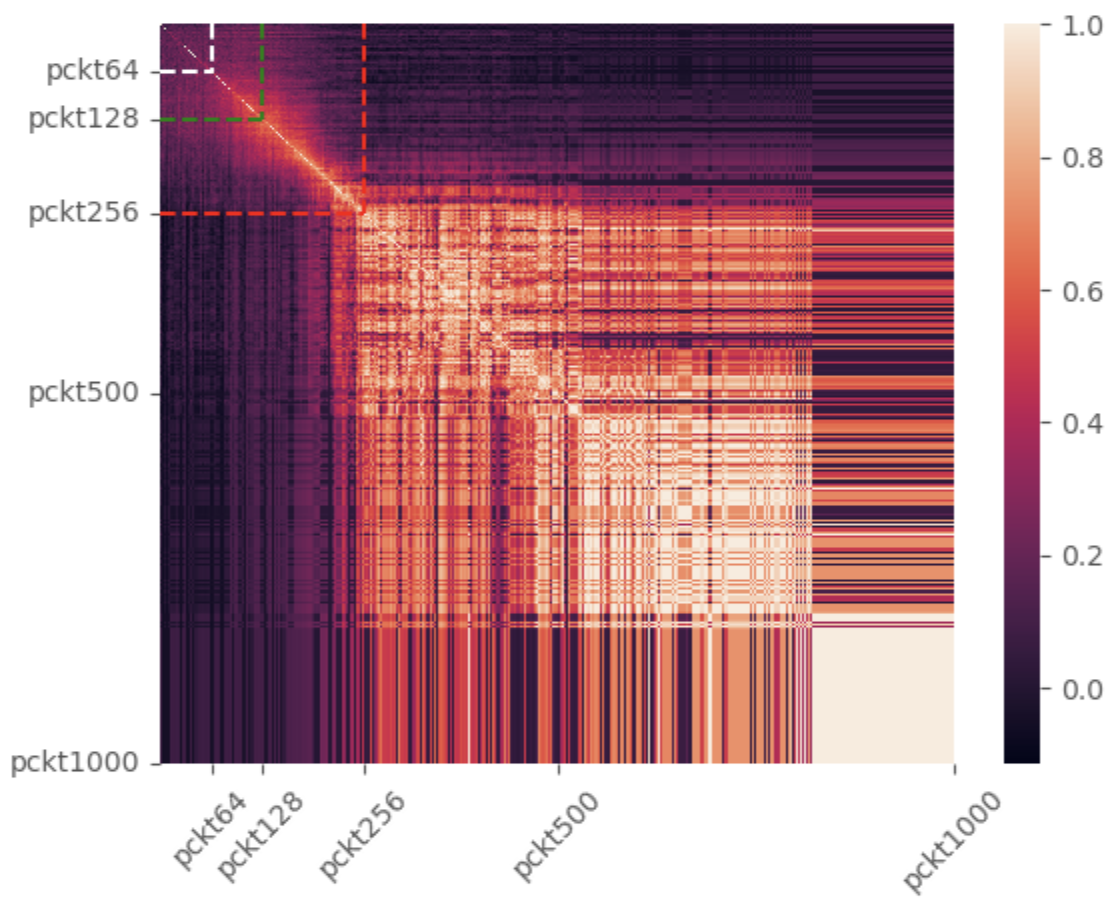}
        \caption{\label{subfig:zo-feat-corr}Correlation heatmap of the features extracted from the ZO dataset.}
    \end{subfigure}
    \caption{\label{fig:zrs-corr}Representative quiet segments (a, c) and pairwise feature correlation heatmaps (b, d) for the WA and ZO datasets, respectively.}
\end{figure*}

\begin{table*}[htbp!]
    \begin{center}
    \caption{Effect of the number of selected features on model performance under different stationarity-based feature selection settings. The best result in each category is highlighted in light gray.}
    \label{tab:num-feats-results}
    
        \begin{adjustbox}{max width=\textwidth}
            \begin{tabular}{| c | c | c | c | c | c | c | c | c | c | c |}
            \hline
                Dataset & Features                       & \makecell[c]{Number of\\Feature} & \makecell[c]{Stationarity\\(1\% / 5\% / 10\%)} & MAPE $\downarrow$             & MSE $\downarrow$      & RMSE $\downarrow$    & MAE $\downarrow$     & LCC $\uparrow$     & SROCC $\uparrow$   & $R^2$ $\uparrow$ \\ 
                \hline
                    \multirow{9}*{\makecell[c]{WhatsApp\\(BRISQUE)}} & \multirow{3}*{PCKT + PIAT} & 175               & 0.1 / 0.2 / 0.5                 & $ 8.77\pm0.131 $  & 81.1717 & 9.0095 & 5.3186 & 0.7881 & 0.7531 & 0.6004\\
                \cline{3-11}
                                               & & 350               & 0.2 / 0.3 / 0.5                 & $ 7.44\pm0.100 $ & 47.1789 & 6.8687 & 4.4818 & 0.8764 & 0.8190 & 0.7679\\
                \cline{3-11}
                                               & & 700               & 0.6 / 1.2 / 2.3                 & $ 7.30\pm0.096 $ & 47.2181 & 6.8715 & 4.4830 & 0.8781 & 0.8202 & 0.7677\\
                \clineB{2-11}{2}
                    & \multirow{3}*{PCKT}        & 175               & 74.9 / 85.6 / 91.5              & $ 9.74\pm0.140 $ & 82.3792 & 9.0763 & 5.6864 & 0.7753 & 0.7455 & 0.5948\\
                \cline{3-11}
                                               & & 350               & 27.8 / 29.6 / 30.5              & $ 8.24\pm0.115 $ & 58.9994 & 7.6811 & 4.8988 & 0.8444 & 0.8037 & 0.7098\\
                \cline{3-11}
                                               & & 700               & 0.3 / 0.5 / 0.7                 & \colorbox{lightgray}{$ 7.16\pm0.095 $} & \colorbox{lightgray}{44.9974} & \colorbox{lightgray}{6.7080} & \colorbox{lightgray}{4.4030} & \colorbox{lightgray}{0.8837} & \colorbox{lightgray}{0.8253} & \colorbox{lightgray}{0.7787}\\
                \clineB{2-11}{2}
                    & \multirow{3}*{PIAT}        & 175               & 97.2 / 98.4 / 98.8              & $ 9.49\pm0.122 $ & 74.8505 & 8.6516 & 5.7847 & 0.7970 & 0.7451 & 0.6318\\
                \cline{3-11}
                                               & & 350               & 36.6 / 46.4 / 53.9              & $ 9.09\pm0.115 $ & 64.1404 & 8.0088 & 5.4746 & 0.8277 & 0.7611 & 0.6845\\
                \cline{3-11}
                                               & & 700               & 4.9 / 8.0 / 12.7                & $ 8.52\pm0.109 $ & 61.2721 & 7.8277 & 5.2234 & 0.8396 & 0.7718 & 0.6986\\
                \hline
                \hline
                    \multirow{9}*{\makecell[c]{Zoom\\(MOS)}} & \multirow{3}*{PCKT + PIAT} & 64                & 4 / 6.8 / 10                & $ 9.39\pm0.394 $ & 0.1417 & 0.3764 & 0.2857 & 0.4754 & 0.5045 & 0.2188 \\
                \cline{3-11}
                                               & & 128               & 4.4 / 6.4 / 9.2             & $ 9.38\pm0.399 $ & 0.1382 & 0.3718 & 0.2818 & 0.4918 & 0.5204 & 0.2378 \\
                \cline{3-11}
                                               & & 256               & 9.4 / 19.4 / 28.9           & $ 9.22\pm0.375 $ & 0.1320 & 0.3633 & 0.2780 & 0.5224 & 0.5396 & 0.2723\\
                \clineB{2-11}{2}
                   & \multirow{3}*{PCKT}        & 64                & 85.8 / 91.2 / 92.6          & $ 10.73\pm0.412 $ & 0.1612 & 0.4015 & 0.3213 & 0.3533 & 0.3653 & 0.1113\\
                \cline{3-11}
                                               & & 128               & 59.3 / 63.5 / 65.5         & $ 9.44\pm0.382 $ & 0.1383 & 0.3718 & 0.2859 & 0.4894 & 0.4845 & 0.2376 \\
                \cline{3-11}
                                               & & 256               & 9.2 / 14.0 / 17.6           & \colorbox{lightgray}{$ 9.04\pm0.372 $} & \colorbox{lightgray}{0.1339} & \colorbox{lightgray}{0.3660} & \colorbox{lightgray}{0.2764} & \colorbox{lightgray}{0.5201} & \colorbox{lightgray}{0.5219} & \colorbox{lightgray}{0.2615}\\
                \clineB{2-11}{2}
                    & \multirow{3}*{PIAT}      & 64                & 87.0 / 90.6 / 91.6          & $ 11.70\pm0.454 $ & 0.1851 & 0.4303 & 0.3477 & -0.0047 & 0.0488 & -0.0210\\
                \cline{3-11}
                                               & & 128               & 58.9 / 63.9 / 67.1          & $ 9.74\pm0.369 $ & 0.1418 & 0.3765 & 0.2958 & 0.4705 & 0.3884 & 0.2181\\
                \cline{3-11}
                                               & & 256               & 12.4 / 18.8 / 24.6          & $ 9.71\pm0.369 $ & 0.1419 & 0.3767 & 0.2953 & 0.4684 & 0.4355 & 0.2174 \\
                \hline
            \end{tabular}
            \end{adjustbox}
    \end{center}
\end{table*}

The first design decision concerned the number of input features to be used by the model. Although each traffic sample was initially represented by up to 1500 packet measurements, exploratory analysis revealed that a substantial fraction of the samples contained significantly fewer valid observations. In particular, many samples in the WA dataset terminated at approximately packet index 800 (\cref{subfig:wa-zrs-hist}), while samples in the ZO dataset often terminated at approximately packet index 350 (\cref{subfig:zo-zrs-hist}). Beyond these points, the remaining feature values were predominantly zero-padded.

This observation is further supported by the feature correlation analysis. As shown in \cref{subfig:wa-feat-corr} and \cref{subfig:zo-feat-corr}, features corresponding to packet indices greater than approximately 800 in the WA dataset and 350 in the ZO dataset exhibit very high mutual correlation, indicating limited additional information content.

To evaluate the impact of the number of input features on prediction performance, we trained the proposed model using several feature lengths. For the WA dataset, we considered input lengths of 128, 350, and 700 features, while for the ZO dataset we evaluated 64, 128, and 256 features. These values were selected to represent regions before, near, and beyond the dominant peaks observed in the distributions shown in \cref{subfig:wa-zrs-hist} and \cref{subfig:zo-zrs-hist}, respectively. The resulting performance is summarized in \cref{tab:num-feats-results}.

The results indicate that models trained using PCKT features consistently outperform those based on PIAT features alone, as well as models utilizing the combined PCKT+PIAT feature set. This observation suggests that packet sizes contain substantially more information relevant to QoE prediction than packet inter-arrival times in the considered datasets. Consequently, unless stated otherwise, all subsequent experiments and analyses are performed using PCKT features only.



\begin{figure*}[htbp!]
    \centering
    \begin{subfigure}[t]{0.45\linewidth}
        \centering 
        \includegraphics[width=1.0\linewidth]{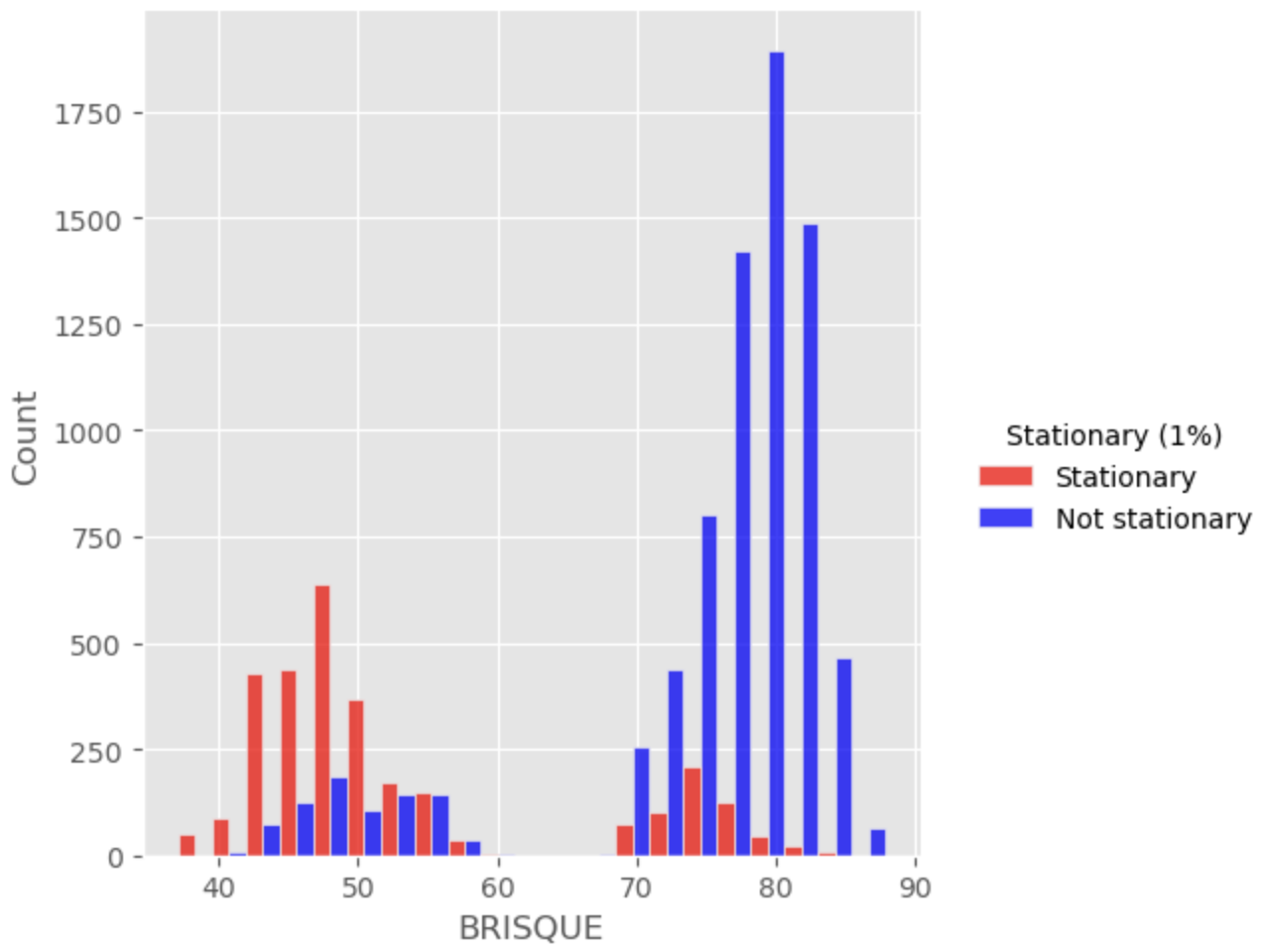}
        \caption{\label{subfig:wa-pckt-stat-nonstat-dist}Distribution of stationary and non-stationary signals in the WA dataset at a significance level of 1\%.}
    \end{subfigure}
    \begin{subfigure}[t]{0.45\linewidth}
        \centering 
        \includegraphics[width=1.0\linewidth]{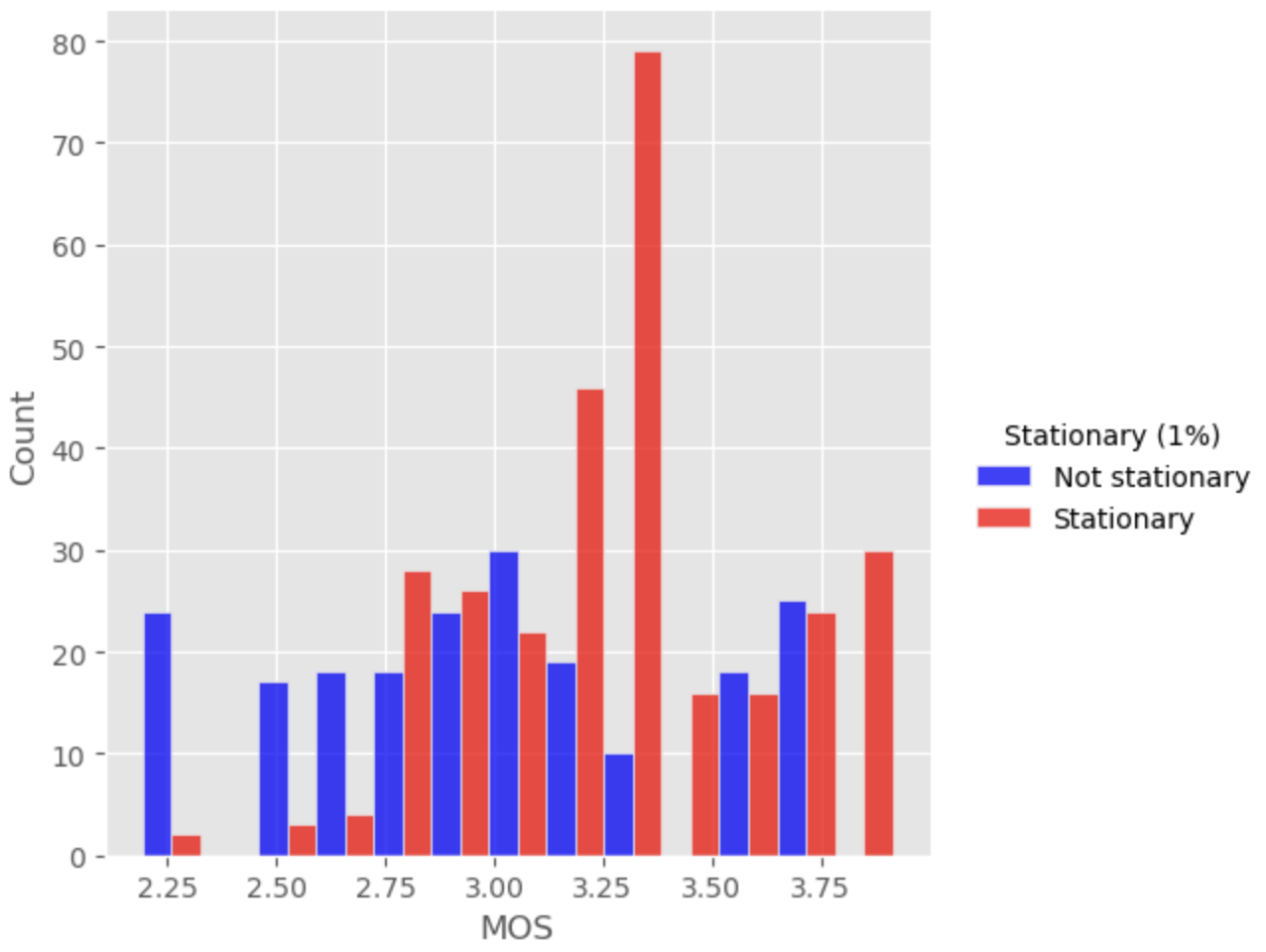}
        \caption{\label{subfig:zo-pckt-stat-nonstat-dist}Distribution of stationary and non-stationary signals in the ZO dataset at a significance level of 1\%.}
    \end{subfigure}
    \caption{\label{fig:stat-nonstat-hist}Distribution of stationary and non-stationary signals in the WA and ZO datasets, determined using the stationarity test described in \cref{sub:stationarity} at a significance level of 1\%.}
\end{figure*}

The results suggest a strong relationship between prediction performance and the proportion of stationary and non-stationary samples in both datasets. In particular, the best MAEP values were obtained for feature configurations containing the lowest proportion of stationary samples (\cref{tab:num-feats-results}).

One possible explanation is that stationary signals typically correspond to stable network conditions, where traffic characteristics exhibit relatively little variation and are therefore less informative for QoE prediction. In contrast, non-stationary signals may reflect changes in network conditions, such as bandwidth fluctuations, packet loss, congestion, or reconnection events, all of which are directly related to QoE degradation and may therefore be easier for the model to identify and exploit.

This hypothesis is further supported by the results presented in \cref{subfig:wa-pckt-stat-nonstat-dist}, where prediction performance improves as the proportion of non-stationary samples increases. Although a similar trend is not clearly visible in the ZO dataset (\cref{subfig:zo-pckt-stat-nonstat-dist}), this discrepancy may be attributed to the substantially smaller size of the ZO dataset, which contains only 500 samples and is therefore more susceptible to statistical variability.

%% file: sections/experiments.tex
\section{Experiments}
\label{sec:experiments}
In this section, we describe the experiments conducted during the study to validate the proposed model's performance and compare it with models commonly used in this domain. 

In all the experiments, we used a 5-fold cross-validation (CV) technique to test for robustness of the results. were 

\subsection{Evaluation Metrics}
\label{sub:metrics}
To evaluate the performance of the proposed approach, we employed a set of standard regression metrics commonly used for predictive modeling, as described in this subsection. 

\subsubsection{Mean Absolute Percentage Error (MAPE)}
\label{subsub:mape}
The primary evaluation metric was the Mean Absolute Percentage Error (MAPE), defined as

\begin{equation}
    MAPE(\textbf{y}, \hat{\textbf{y}}) = \frac{100}{B} \sum_{i=0}^{B}|1-\frac{\hat{y}_i}{y_i}|
    \label{eq:maep}
\end{equation}

\noindent where $M$ represents the number of data points in the vector, $\mathbf{y}$ and $\hat{\mathbf{y}}$ are the true and predicted signals respectively. This metric may be interpreted as the model's scaleless prediction accuracy, expressed as a percentage of error relative to the true label. 

\subsubsection{Mean Squared Error (MSE)}
\label{subsub:mse}
This metric evaluates the mean squared distance between the true values and the predicted ones by

\begin{equation}
    MSE(\mathbf{y}, \hat{\mathbf{y}}) = \frac{1}{N} \sum_{i=1}^{N} (y_i - \hat{y}_i)^2
    \label{eq:mse}
\end{equation}
\noindent where $y_i$ is the true value, while $\hat{y}_i$ is the predicted value for some observation $i$. Since the MSE represents distance, lower values are preferred because they correspond to better predictive performance.

\subsubsection{Root Mean Squared Error (RMSE)}
\begin{equation}
    RMSE(\mathbf{y}, \hat{\mathbf{y}}) = \sqrt{MSE(\mathbf{y}, \hat{\mathbf{y}})}
    \label{eq:rmse}
\end{equation}
As MSE is squared, to produce errors in the same units as the predicted values, RMSE is used.

\subsubsection{Mean Absolute Error (MAE)}
\begin{equation}
    MAE(\mathbf{y}, \hat{\mathbf{y}}) = \frac{1}{N} \sum_{i=1}^{N} |y_i - \hat{y}_i|
    \label{eq:mae}
\end{equation}
Because MSE and RMSE are squared, they may exhibit error skewness toward large errors, whereas MAE treats all errors equally, making it more robust to outliers. Additionally, as MAE uses the same units as the data, it has better interpretability.

\subsubsection{Linear Correlation Coefficient (LCC)}
\label{subsub:lcc}

LCC, also known as Pearson's correlation coefficient, measures the strength and direction of the linear relationship between two variables. In our case, these variables correspond to the ground-truth QoE labels, denoted by $\mathbf{y}$, and the predicted QoE values, denoted by $\hat{\mathbf{y}}$. LCC quantifies the degree of linear association by normalizing the covariance of the two variables by the product of their standard deviations:

\begin{equation}
    LCC(\mathbf{y}, \hat{\mathbf{y}}) =
    \frac{Cov_{\mathbf{y}, \hat{\mathbf{y}}}}
    {\sigma_{\mathbf{y}}\sigma_{\hat{\mathbf{y}}}}
    \label{eq:lcc}
\end{equation}

\noindent where the covariance between the ground-truth and predicted values is defined as

\begin{equation}
    Cov_{\mathbf{y}, \hat{\mathbf{y}}} =
    \frac{1}{M}\sum_{i=1}^{M}
    (y_i - \mu_{\mathbf{y}})
    (\hat{y}_i - \mu_{\hat{\mathbf{y}}})
    \label{eq:cov}
\end{equation}

\noindent and the corresponding sample means are given by

\begin{equation}
    \mu_{\mathbf{y}} =
    \frac{1}{M}\sum_{i=1}^{M} y_i
    \label{eq:mu}
\end{equation}

\begin{equation}
    \mu_{\hat{\mathbf{y}}} =
    \frac{1}{M}\sum_{i=1}^{M} \hat{y}_i
    \label{eq:mu_hat}
\end{equation}

Intuitively, the covariance defined in \cref{eq:cov} measures the extent to which the ground-truth values $\mathbf{y}$ and the predicted values $\hat{\mathbf{y}}$ vary together. Positive covariance indicates that both variables tend to increase or decrease together, whereas negative covariance indicates that they tend to move in opposite directions. By normalizing the covariance with the standard deviations of the two variables, LCC provides a scale-independent measure of linear association.

LCC values lie in the range $[-1, 1]$. A value of $1$ indicates a perfect positive linear relationship, whereas a value of $-1$ corresponds to a perfect negative linear relationship. Values close to $0$ indicate little or no linear correlation between the variables. In the context of QoE prediction, higher positive LCC values indicate better agreement between the predicted and ground-truth QoE scores, while negative values suggest an undesirable inverse relationship.

\subsubsection{Spearman's Rank-Order Correlation Coefficient (SROCC)}

While the LCC \cref{subsub:lcc} measures the strength of the linear relationship between the ground-truth and predicted QoE values, \(\mathbf{y}\) and \(\hat{\mathbf{y}}\), respectively, the SROCC evaluates the degree of monotonic association between these variables. Consequently, SROCC is particularly useful when the relationship between the variables is non-linear.

SROCC is defined as

\begin{equation}
    SROCC(\mathbf{y}, \hat{\mathbf{y}}) =
    1 - \frac{6 \sum_{i=1}^{N} \Delta R(i)^2}
    {N(N^2 - 1)}
    \label{eq:srocc_coeff}
\end{equation}

\noindent where \(N\) denotes the number of samples and \(\Delta R(i)\) represents the difference between the ranks assigned to the \(i\)-th ground-truth and predicted values. Specifically, each sample is assigned a rank \(R_{\mathbf{y}}(i)\) and \(R_{\hat{\mathbf{y}}}(i)\) according to its position in the ordered sets \(\mathbf{y}\) and \(\hat{\mathbf{y}}\), respectively. The rank difference is then computed as

\begin{equation}
    \Delta R(i) =
    R_{\mathbf{y}}(i) -
    R_{\hat{\mathbf{y}}}(i)
    \label{eq:srocc_delta}
\end{equation}

The squared rank differences,

\begin{equation}
    \Delta R(i)^2
    \label{eq:srocc_delta_sq}
\end{equation}

\noindent penalize larger discrepancies between the ranks more heavily than smaller ones. The overall disagreement between the rankings of the ground-truth and predicted values is quantified by the sum

\begin{equation}
    \sum_{i=1}^{N} \Delta R(i)^2
    \label{eq:srocc_sum_deltas}
\end{equation}

The normalization term \(N(N^2-1)\) in \cref{eq:srocc_coeff} ensures that the resulting coefficient is bounded within the interval \([-1,1]\).

SROCC values range from \(-1\) to \(1\). A value of \(1\) indicates perfect agreement between the rankings of the ground-truth and predicted values, while a value of \(-1\) corresponds to a perfect inverse ranking. A value close to \(0\) indicates little or no monotonic relationship between the variables.

SROCC is particularly relevant for QoE prediction tasks, as both objective and subjective QoE measures often exhibit non-linear relationships with the underlying network and service characteristics. By evaluating rank consistency rather than strict linear correspondence, SROCC provides a robust measure of predictive performance. Furthermore, its bounded range and intuitive interpretation facilitate straightforward comparison between different models and experimental settings.

\subsubsection{Coefficient of Determination ($R^2$)}

The coefficient of determination, denoted by $R^2$, is a goodness-of-fit metric that quantifies the proportion of variance in the ground-truth values $\mathbf{y}$ that is explained by the predicted values $\hat{\mathbf{y}}$. It is defined as

\begin{equation}
R^2(\mathbf{y}, \hat{\mathbf{y}}) =
1 -
\frac{\sum_{i=1}^{N} (y_i - \hat{y}_i)^2}
{\sum_{i=1}^{N} (y_i - \mu_{\mathbf{y}})^2}
\label{eq:r_sq}
\end{equation}

\noindent where $y_i$ and $\hat{y}_i$ denote the ground-truth and predicted values of the $i$-th sample, respectively, and $\mu_{\mathbf{y}}$ is defined as in \cref{eq:mu}, and represents the mean of the ground-truth values.

The numerator in \cref{eq:r_sq} represents the residual sum of squares (RSS), which measures the unexplained variation in the data, while the denominator corresponds to the total sum of squares (TSS), which measures the overall variation in the ground-truth values. Consequently, $R^2$ quantifies the fraction of the total variance that is captured by the model.

In general, $R^2$ values closer to $1$ indicate that the model explains a larger proportion of the data's variability and therefore exhibits better predictive performance. A value of $R^2 = 0$ indicates that the model performs no better than predicting the mean of the ground-truth values. Negative $R^2$ values are also possible and indicate that the model performs worse than this baseline predictor.

\subsection{Pipeline Description}
\label{subsub:training}
This subsection describes the training and testing pipeline used in our experiments and discusses the key algorithmic decisions underlying its design.

\subsubsection{Training Procedure}
The models were trained for 100 epochs. To obtain a robust estimate of predictive performance, we employed 5-fold cross-validation in all experiments. In each fold, 80\% of the dataset was used for training and the remaining 20\% was reserved for testing. The test partitions were mutually exclusive, ensuring that each sample appeared in the test set exactly once across the five evaluation runs. The reported results correspond to the average performance across all folds.

\subsubsection{Pretrained Head Model Weights}
\begin{table*}[htbp!]
    \begin{center}
    \caption{Performance of the proposed qCNN model obtained by freezing different numbers of residual basic blocks (BBlocks) in the HEAD module (\cref{fig:qcnn_arch} (b)).}
    \label{tab:layers-freeze-results}
    
        \begin{adjustbox}{max width=\textwidth}
            \begin{tabular}{| c | c | c | c | c | c | c | c |}
            \hline
                     \makecell[c]{Number of\\Frozen Layers}          & MAPE $\downarrow$             & MSE $\downarrow$      & RMSE $\downarrow$    & MAE $\downarrow$     & LCC $\uparrow$     & SROCC $\uparrow$   & $R^2$ $\uparrow$ \\ 
                \hline
                      0             & $ 7.70\pm0.103 $ & 50.8800 & 7.1330 & 4.6826 & 0.8660 & 0.8141 & 0.7497\\
                \hline
                                                        1             & $ 7.81\pm0.108 $ & 52.8140 & 7.2673 & 4.6764 & 0.8607 & 0.8096 & 0.7402 \\
                \hline
                                                        2                 & \colorbox{lightgray}{$ 7.16\pm0.095 $}   & \colorbox{lightgray}{44.9974}  & \colorbox{lightgray}{6.7080} & \colorbox{lightgray}{4.4030} & \colorbox{lightgray}{0.8837} & \colorbox{lightgray}{0.8253} & \colorbox{lightgray}{0.7787}\\
                \hline
                                                        3                 & $ 7.26\pm0.099 $ & 46.3556 & 6.8085 & 4.4181 & 0.8795 & 0.8110 & 0.7720\\
                \hline
                                                        4  & $ 7.34\pm0.097 $ & 46.3951 & 6.8114 & 4.4858 & 0.8791 & 0.8138 & 0.7718\\
                \hline
            \end{tabular}
            \end{adjustbox}
    \end{center}
\end{table*}

Instead of training the HEAD module from randomly initialized weights, we initialized it using weights pre-trained on the ImageNet1k dataset. Transfer learning from ImageNet has been shown to improve performance across a wide range of tasks, including domains that differ substantially from natural image classification, due to the ability of the early convolutional layers to capture generic low-level patterns such as edges, contours, and simple geometric structures \cite{kornblith2019better, ebbehoj2022transfer}. Consequently, the use of pre-trained weights can accelerate convergence and improve generalization, particularly when the available training data are limited.

The use of ImageNet-pretrained weights is additionally motivated by prior studies demonstrating the effectiveness of convolutional feature extractors on non-natural image representations. In particular, FlowPic \cite{shapira2019flowpic} successfully applied CNNs to image-like representations of network traffic, while Segal et al. \cite{segal2023using} employed ImageNet-pretrained EfficientNet models on TSSCI representations derived from human motion data. Together with our previous findings in \cite{sidorov2024revisiting}, these results suggest that convolutional filters pretrained on large-scale image datasets can provide useful initialization even when the input originates from a different domain, provided that it is represented in a structured two-dimensional form.

To adapt the model to the traffic analysis domain, only the first two residual basic blocks (BBlocks) of the HEAD module (\cref{fig:qcnn_arch} (b)) were kept frozen, while the remaining layers were allowed to update during training. This strategy preserves the generic low-level representations learned from ImageNet while enabling the higher-level features to specialize for QoE prediction.

To validate this design choice, we conducted an ablation study in which different numbers of BBlocks were frozen during training. The results, presented in \cref{tab:layers-freeze-results}, demonstrate that freezing the first two BBlocks yields the best overall predictive performance.

\subsubsection{Saw-LR Learning Rate (LR) Schedule}
\label{subsub:saw-lr}

\begin{figure}[htbp!]
    \centering
    \includegraphics[width=1.0\linewidth]{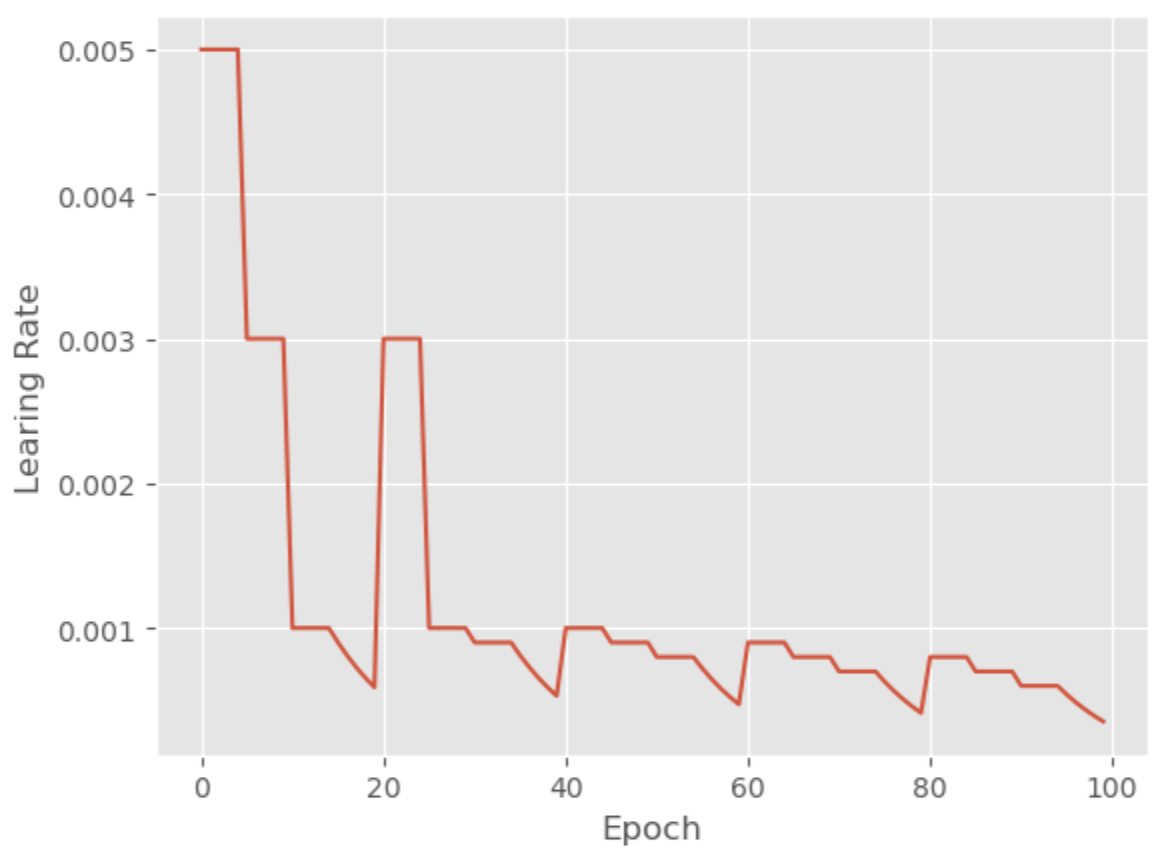}
    \caption{\label{fig:saw-lr}Proposed version of a cyclical LR used in this work. The LR value is gradually decreased over the course of training, with spikes to the high learning rates to escape local minima traps.}
\end{figure}

The proposed learning rate schedule is inspired by the cyclical learning rate strategy introduced by Smith \cite{smith2017cyclical}, which was designed to improve optimization by periodically increasing the learning rate and thereby facilitating exploration of the parameter space. We refer to the resulting schedule as the \textit{Saw-LR} schedule due to its characteristic sawtooth-shaped profile, as illustrated in \cref{fig:saw-lr}.

\begin{algorithm}
\caption{A saw-lr learning rate schedule algorithm}\label{alg:saw-lr}
\textbf{Input} $lr_{init}$\\
\textbf{Output} $lr$
\begin{algorithmic}[1]
\State $P \gets (\lceil\frac{Epoch}{5}\rceil \bmod 5) + 1$
\If{$P = 1$}
    \State $lr \gets lr_{init}$
\ElsIf{$P = 2$}
    \State $lr_{init} \gets lr$
    \If{$lr > 1 \times 10^{-3}$}
        \State $lr \gets lr_{init} - 2^{P - 1} \times 10^{-3}$
    \Else
        \State $lr \gets lr_{init} - 1 \times 10^{-4}$
    \EndIf
\ElsIf{$P = 3$}
    \If{$lr > 1 \times 10^{-3}$}
        \State $lr \gets lr_{init} - 2^{P - 1} \times 10^{-3}$
    \Else
        \State $lr \gets lr_{init} - 1 \times 10^{-4}$
    \EndIf
\Else
    \If{$lr > 1 \times 10^{-5}$}
        \State $lr \gets lr \times 0.9^{Epoch \bmod 5 + 1}$
    \EndIf
\EndIf
\RETURN{} $lr$
\end{algorithmic}
\end{algorithm}

The schedule operates in rounds ($R$), where each round consists of four phases ($P$), and each phase spans five training epochs. To encourage rapid initial convergence, training begins with a learning rate of $5 \times 10^{-3}$, which is maintained throughout the first phase ($P=1$). During the second phase ($P=2$), the learning rate is reduced by $2 \times 10^{-3}$ and kept constant for the subsequent five epochs. In the third phase ($P=3$), it is further reduced by an additional $2 \times 10^{-3}$ and maintained for another five epochs.

During the fourth phase ($P=4$), the learning rate is progressively reduced to facilitate stable convergence. Specifically, at epoch (E), the learning rate is updated according to

\begin{equation}
    LR \leftarrow LR \cdot 0.9^{(E \bmod 5)+1},
    \label{eq:lr-5-phase}    
\end{equation}

\noindent provided that $LR > 10^{-5}$. Once the learning rate reaches this threshold, it is reset to its previous value to avoid excessively small parameter updates that could slow or stall the optimization process.

At the beginning of each new round, the learning rate is restored to the value used during phase ($P=2$) of the preceding round. This controlled increase enables the optimizer to move away from the current solution and continue exploring alternative regions of the loss landscape, thereby reducing the likelihood of premature convergence. The resulting learning rate schedule is illustrated in \cref{fig:saw-lr}, while the full procedure is provided in \cref{alg:saw-lr}.

\subsubsection{Regularization}
\label{subsub:regularization}
To mitigate overfitting, we employed Dropout \cite{srivastava2014dropout}, a regularization technique that randomly deactivates an arbitrary number of neurons during model training. During the first 10 training epochs, dropout was disabled ($p_{drop}=0$) to facilitate rapid convergence and allow the model to learn stable feature representations. Once the training process stabilized, dropout was enabled with a probability of $p_{drop}=0.01$, meaning that approximately 1\% of the neurons in the FC layer were randomly deactivated during each training iteration. Given the relatively small size of the FC layer, this corresponds to approximately five neurons being dropped on average. The selected dropout rate was determined empirically and provided the best trade-off between regularization and predictive performance on the validation set. The dropout probability remained constant for the remainder of the training process.

\subsubsection{Augmentation}
For data augmentation, we employed two simple techniques: additive Gaussian noise and input-vector shuffling. Although packet sizes reflect latent properties of the underlying communication protocols and adaptive streaming algorithms, they are not deterministic and exhibit natural variability. Consequently, samples associated with similar QoE levels may still differ in their low-level traffic characteristics. To improve the robustness and generalization capability of the model, two augmentation operations were applied independently with probability ($p_{\mathrm{aug}} = 0.2$).

The first augmentation operation consisted of adding Gaussian noise sampled from ($N(0,1)$) to the input feature vector. This perturbation reduces the model's sensitivity to minor fluctuations in the measured traffic characteristics and encourages the learning of more stable feature representations.

The second augmentation operation consisted of randomly shuffling the elements of the input vector. Although this transformation does not correspond to a realistic traffic-generation process, it acts as a regularization mechanism by preventing the model from relying excessively on a particular arrangement of measurements. In the case of UDP traffic, where packet ordering is not guaranteed by the transport protocol, moderate changes in packet ordering may naturally occur. For TCP traffic, shuffling intentionally disrupts temporal dependencies and therefore produces samples corresponding to degraded or atypical communication conditions. As a result, the augmentation encourages the model to learn traffic patterns that are robust to temporal perturbations while preserving sensitivity to traffic characteristics associated with QoE degradation.

\subsubsection{Optimizer}
The qCNN model was trained using the Adam optimizer \cite{kingma2014adam} with its default hyperparameter settings. Adam combines adaptive learning-rate estimation with momentum-based optimization, enabling efficient and stable training of deep neural networks. Owing to its strong empirical performance across a wide range of machine learning tasks, it has become one of the most widely adopted optimization algorithms in deep learning and was therefore selected as the optimization method in all experiments.

\subsubsection{Loss Function}
Following common practice in regression-based QoE prediction, we employed the MSE (\cref{subsub:mse}) as the training objective. The loss function is defined as
\begin{equation}
    \mathcal{L}(\mathbf{y}, \hat{\mathbf{y}}) = \frac{1}{B} \sum_{i=1}^{B} (y_i - \hat{y}_i)^2
\end{equation}
\noindent where $y_i$ and $\hat{y}_i$ denote the ground-truth and predicted QoE values for the $i$-th sample, respectively, and $B$ is the mini-batch size. By penalizing the squared prediction error, the MSE loss encourages the model to minimize large deviations between the predicted and target QoE values, thereby improving regression accuracy during training.

\subsection{Reference Models}
\label{sub}

In this subsection, we describe the reference models used to evaluate the performance of the proposed qCNN architecture. All reference models were trained and evaluated using the same experimental protocol as qCNN, as described in \cref{subsub}, ensuring a fair and consistent comparison.

\subsubsection{Classical Machine Learning (ML)-Based Models}

The classical machine learning baselines included Support Vector Regression (SVR) with a radial basis function (RBF) kernel \cite{cortes1995support}, Random Forest (RF) \cite{ho1995random}, and several ensemble learning methods, including AdaBoost, XGBoost \cite{chen2016xgboost}, and CatBoost \cite{prokhorenkova2018catboost}. These models were selected because they represent widely adopted regression techniques with diverse learning paradigms, ranging from kernel-based methods to bagging- and boosting-based ensembles. Furthermore, they have consistently demonstrated strong performance across a broad range of prediction tasks and therefore constitute meaningful non-deep-learning baselines.

\subsubsection{Deep Learning (DL)-Based Models}

To assess the effectiveness of the proposed architecture relative to existing deep learning approaches, we selected three representative models that have previously demonstrated competitive performance in QoE prediction tasks: CNNQoE \cite{duc2020convolutional}, TCN \cite{bai2018empirical}, and LSTM-QoE \cite{eswara2019streaming}. These architectures represent three distinct paradigms for sequence modeling, namely convolutional, temporal convolutional, and recurrent neural networks, respectively.

Since the original source code was not publicly available, each model was reimplemented based on the descriptions provided in the corresponding publications. The architectures, training procedures, and hyperparameter settings were reproduced as closely as possible to those reported by the original authors in order to ensure a faithful comparison.

%% file: sections/ablation.tex
\section{Ablation Study}
\label{sec}
In this section, we evaluate the contribution of the main architectural components and algorithmic design choices of the proposed qCNN model through an ablation study. The goal is to quantify the impact of each component on prediction performance and to justify its inclusion in the final architecture.

\begin{table*}[htbp!]
    \begin{center}
    \caption{Results of the ablation study conducted on the WA and ZO datasets used in this work.}
    \label{tab:ablation-results}
    
        \begin{adjustbox}{max width=\textwidth}
            \begin{tabular}{| c | c | c | c | c | c | c | c | c |}
            \hline
                     Dataset                           & Configuration          & MAPE $\downarrow$             & MSE $\downarrow$      & RMSE $\downarrow$    & MAE $\downarrow$     & LCC $\uparrow$     & SROCC $\uparrow$   & $R^2$ $\uparrow$ \\ 
                \hline
                     \multirow{8}*{\makecell[c]{WA\\(BRISQUE)}} & Random weight init.             & $ 7.47\pm0.100 $ & 48.1394 & 6.9383 & 4.5141 & 0.8739 & 0.8165 & 0.7632\\
                \cline{2-9}
                                                       & Freeze EMBD             & $ 7.57\pm0.103 $ & 48.8112 & 6.9865 & 4.5146 & 0.8727 & 0.8132 & 0.7599 \\
                \cline{2-9}
                                                       & No REG                 & $ 7.98\pm0.101 $    & 51.8401  & 7.2000  & 4.8545  & 0.8635  & 0.8022  & 0.7450\\
                \cline{2-9}
                                                       & No CNN                 & $ 20.80\pm0.183 $ & 201.4344 & 14.1928 & 11.9917 & 0.2557 & 0.2408 & 0.0092\\
                \cline{2-9}
                                                       & LR = $1 \cdot 10^{-3}$ & $ 8.01\pm0.106 $  & 54.0367  & 7.3510  & 4.8242  & 0.8570  & 0.7987  & 0.7342\\
                \cline{2-9}
                                                       & LR = $5 \cdot 10^{-3}$ & $ 8.18\pm0.106 $  & 53.5594  & 7.3184  & 4.8962  & 0.8619  & 0.7934  & 0.7366\\
                \cline{2-9}
                                                       & LR = $1 \cdot 10^{-4}$ & $ 7.88\pm0.107 $  & 52.3106  & 7.2326  & 4.6806  & 0.8649  & 0.8152  & 0.7427\\
                \hline
                \hline
                     \multirow{8}*{\makecell[c]{ZO\\(MOS)}}         & Random weight init.             & $ 9.10\pm0.375 $  & 0.1309   & 0.3618  & 0.2759  & 0.5307  & 0.5410  & 0.2782\\
                \cline{2-9}
                                                       & Freeze EMBD             & $ 9.54\pm0.403 $  &  0.1451  &  0.3809 & 0.2879  &  0.4522 &  0.4738 & 0.1999\\
                \cline{2-9}
                                                       & No REG                 & $ 9.56\pm0.398 $  & 0.1483   & 0.3851  & 0.2900  & 0.4431  & 0.4973  & 0.1822\\
                \cline{2-9}
                                                       & No CNN                 & $ 11.62\pm0.468 $ & 0.1813   & 0.4258  & 0.3415  & 0.0725  & 0.0559  & 0.0003\\
                \cline{2-9}
                                                       & LR = $1 \cdot 10^{-3}$ & $ 10.41\pm0.413 $ & 0.1583   & 0.3979  & 0.3113  & 0.4270  & 0.4667  & 0.1270\\
                \cline{2-9}
                                                       & LR = $5 \cdot 10^{-3}$ & $ 9.75\pm0.385 $  & 0.1395   & 0.3735  & 0.2928  & 0.4865  & 0.5034  & 0.2307\\
                \cline{2-9}
                                                       & LR = $1 \cdot 10^{-4}$ & $ 9.88\pm0.407 $  & 0.1551   & 0.3938  & 0.3000  & 0.4379  & 0.4644  & 0.1449\\
                \hline
            \end{tabular}
            \end{adjustbox}
    \end{center}
\end{table*}

The results of the ablation study are summarized in \cref{tab:ablation-results}. Experiments were conducted on both datasets used in this work, namely WA and ZO, following the training and evaluation procedure described in \cref{subsub:training}. For each dataset, we evaluated several modified versions of the proposed architecture, each obtained by removing or altering a single design component. The resulting performance degradation relative to the full qCNN model provides an estimate of the contribution of the corresponding component.

As shown in \cref{tab:ablation-results}, the CNN HEAD module constitutes the most influential architectural component, as its removal leads to the largest performance degradation across both datasets. This finding highlights the importance of convolutional feature extraction for identifying traffic patterns that are informative for QoE prediction.

We further investigated the impact of the learning-rate schedule by replacing the proposed Saw-LR strategy (\cref{subsub:saw-lr}) with fixed learning rates commonly used in conjunction with the Adam optimizer, namely $1 \times 10^{-3}$, $5 \times 10^{-3}$, and $1 \times 10^{-4}$. In all cases, the resulting models exhibited inferior performance, supporting the effectiveness of the proposed learning-rate schedule.

The \textit{No REG} configuration corresponds to training the model without the regularization techniques described in \cref{subsub:regularization}. The \textit{Freeze EMBD} configuration denotes a model in which the parameters of the EMBD module remain fixed throughout training and are therefore not updated via backpropagation. Finally, the \textit{Random Weight Initialization} configuration refers to training the model entirely from scratch using randomly initialized weights. In this setting, all layers of the CNN HEAD module are trainable, in contrast to the proposed transfer-learning configuration, where weights pretrained on ImageNet-1K are used to initialize the feature extractor while preserving generic low-level representations learned from large-scale image data.

Overall, the ablation study demonstrates that all proposed architectural components and training strategies contribute positively to the final predictive performance. Among the investigated design choices, the CNN HEAD module and the Saw-LR learning-rate schedule provide the largest performance improvements, indicating that both effective feature extraction and optimization play a critical role in the success of the proposed qCNN framework.

%% file: sections/results.tex
\section{Results} 
\label{sec:results}
\begin{table*}[htbp!]
    \begin{center}
    \caption{Prediction performance for BRISQUE and MOS estimation on the WA and ZO datasets.}
    \label{tab:results}
        \begin{adjustbox}{max width=\textwidth}
            \begin{tabular}{| c | c | c | c | c | c | c | c | c |}
            \hline
                Dataset                          & Model                               & MAPE $\downarrow$             & MSE $\downarrow$      & RMSE $\downarrow$    & MAE $\downarrow$     & LCC $\uparrow$     & SROCC $\uparrow$   & $R^2$ $\uparrow$ \\ 
            \hline
               \multirow{9}*{\makecell[c]{WA\\(BRISQUE)}} & Random Forest \cite{ho1995random}                       & $ 9.26\pm0.106 $ & 65.5222 & 8.0946 & 5.7217 & 0.8247 & 0.7615 & 0.6777\\
            \cline{2-9}
                                                 & SVM \cite{cortes1995support}                                & $ 8.27\pm0.117 $ & 62.1868 & 7.8859 & 4.9789 & 0.8350 & 0.7663 & 0.6941\\
            \cline{2-9}
                                                 & AdaBoost \cite{freund1995desicion}                           & $ 15.54\pm0.080 $ & 119.5136 & 10.9322 & 10.1158 & 0.7893 & 0.7645 & 0.4121\\
            \cline{2-9}
                                                 & XGBoost \cite{chen2016xgboost}     & $ 8.65\pm0.104 $ & 60.9172 & 7.8049 & 5.3426 & 0.8381 & 0.7801 & 0.7004\\
            \cline{2-9}
                                                 & CatBoost \cite{prokhorenkova2018catboost} & $ 7.91\pm0.097 $ & 50.1883 & 7.0844 & 4.8441 & 0.8679 & 0.8118 & 0.7531\\
            \cline{2-9}
                                                 & CNNQoE \cite{duc2020convolutional}  & $ 16.36\pm0.171 $  & 149.1192 & 12.2114 & 9.4884 & 0.5240 & 0.4387 & 0.2665\\
            \cline{2-9}
                                                 & TCN \cite{bai2018empirical}  & $ 20.99\pm0.184 $  & 204.8701 & 14.3133 & 12.1122 & -0.0014 & 0.0029 & -0.0077\\
            \cline{2-9}
                                                 & LSTM-QoE \cite{eswara2019streaming} & $ 18.76\pm0.169 $  & 169.6362 & 13.0244 & 10.8470 & 0.4591 & 0.4173 & 0.1656\\
            \cline{2-9}
                                                 & qCNN                                & \colorbox{lightgray}{$ 7.16\pm0.095 $}   & \colorbox{lightgray}{44.9974}  & \colorbox{lightgray}{6.7080} & \colorbox{lightgray}{4.4030} & \colorbox{lightgray}{0.8837} & \colorbox{lightgray}{0.8253} & \colorbox{lightgray}{0.7787}\\
            \hline
            \hline
               \multirow{8}*{\makecell[c]{ZO\\(MOS)}}         & Random Forest  \cite{ho1995random}                 & $ 9.56\pm0.364 $ & \colorbox{lightgray}{0.1286} & \colorbox{lightgray}{0.3586} & 0.2856 & \colorbox{lightgray}{0.5526} & \colorbox{lightgray}{0.5225} & \colorbox{lightgray}{0.2908}\\
            \cline{2-9}
                                                 & SVM \cite{cortes1995support}                                & $ 9.25\pm0.338 $ & 0.1236 & 0.3516 & 0.2801 & 0.5654 & 0.5453 & 0.3183\\
            \cline{2-9}
                                                 & AdaBoost       \cite{freund1995desicion}                      & $ 9.97\pm0.356 $ & 0.1326 & 0.3642 & 0.2976 & 0.5515 & 0.5167 & 0.2687\\
            \cline{2-9}
                                                 & XGBoost \cite{chen2016xgboost}     & $ 9.90\pm0.365 $ & 0.1361 & 0.3689 & 0.2977 & 0.5178 & 0.4999 & 0.2497 \\
            \cline{2-9}
                                                 & CatBoost \cite{prokhorenkova2018catboost} & $ 14.72\pm0.534 $ & 83.8116 & 9.1549 & 7.4659 & 0.4596 & 0.4361 & 0.2023 \\
            \cline{2-9}
                                                 & CNNQoE \cite{duc2020convolutional}  & $11.56\pm0.459$    & 0.1798 & 0.4241 & 0.3413 & 0.1061 & 0.1117 & 0.0083\\
            \cline{2-9}
                                                 & TCN \cite{bai2018empirical}  & $13.57\pm0.479$    & 0.2624 & 0.5123 & 0.4135 & -0.0418 & -0.0303 & -0.4472\\
            \cline{2-9}
                                                 & LSTM-QoE \cite{eswara2019streaming} & $11.04\pm0.445$    & 0.1730 & 0.4159 & 0.3268 & 0.2943 & 0.3418 & 0.0461 \\
            \cline{2-9}
                                                 & qCNN                                & \colorbox{lightgray}{$ 9.04\pm0.372 $}  & 0.1339 & 0.3660 & \colorbox{lightgray}{0.2764} & 0.5201 & 0.5219 & 0.2615 \\ 
            \hline
            \end{tabular}
            \end{adjustbox}
    \end{center}
\end{table*}
The performance of the proposed qCNN model was evaluated and compared with the baseline models considered in this study. The final results are summarized in \cref{tab:results} and demonstrate the effectiveness of the proposed approach for QoE prediction from low-level traffic measurements.

As shown in \cref{tab:results}, qCNN achieves the best overall performance on the WA dataset, consistently outperforming the competing methods across the evaluated metrics. On the ZO dataset, the proposed model attains the lowest MAPE and MAE values, namely 9.04 and 0.2764, respectively. These results indicate that the proposed architecture is capable of effectively capturing traffic characteristics that are informative for QoE prediction across different datasets and target quality metrics.

Notably, qCNN maintains strong predictive performance even on the ZO dataset, which contains only 500 samples. This observation suggests that the proposed architecture exhibits good robustness under limited-data conditions and is able to generalize effectively despite the relatively small training set.

Nevertheless, the relatively small size of the ZO dataset limits the statistical strength of the conclusions that can be drawn from this experiment. Therefore, the MOS prediction results should be interpreted as preliminary evidence of the applicability of the proposed approach to subjective QoE estimation. Validation on larger MOS-annotated datasets remains an important direction for future work.

To provide further insight into the prediction quality, scatter plots comparing the ground-truth and predicted values are presented in \cref{appdx-fig:wa-brisque-error-scatter} and \cref{appdx-fig:zo-mos-error-scatter}. These plots illustrate the relationship between the predicted and target values and enable a more detailed examination of the residual prediction errors.





%% file: sections/conclusions_and_future_work.tex
\section{Conclusion and Future Work}
\label{sec:conclusions}

In this work, we proposed qCNN, a novel convolutional framework for QoE prediction from passive QoS measurements collected from encrypted communication channels. The proposed architecture combines a learnable one-dimensional to two-dimensional embedding module with a CNN-based feature extractor and was designed to capture complex traffic patterns without relying on application-layer information. We additionally introduced a dedicated learning-rate scheduling strategy, termed Saw-LR (\cref{subsub:saw-lr}), and demonstrated through extensive experiments that it consistently improves model performance compared with conventional fixed learning-rate settings.

As part of the study, we analyzed traffic traces collected from two widely used video conferencing (VC) platforms, namely WhatsApp and Zoom, focusing on packet size and packet inter-arrival time characteristics. The extracted features were further investigated from the perspective of signal stationarity when interpreted as time-series data. The results suggest that stationary traffic patterns are generally associated with stable communication conditions and higher QoE levels, whereas increased non-stationarity often coincides with network degradation and reduced user-perceived quality.

The proposed qCNN model was evaluated against both classical machine learning baselines and deep learning architectures previously applied to QoE prediction. Experimental results demonstrated that qCNN consistently achieves superior predictive performance across the seven regression metrics considered in this study (\cref{sub:metrics}) on the WA dataset, confirming the effectiveness of the proposed architecture and training strategy.

Despite these promising results, several limitations remain. Although the proposed method was evaluated on datasets collected from two different VC platforms, the Zoom dataset was relatively small, containing only 500 samples. While qCNN achieved the best MAPE and MAE values for MOS prediction on this dataset, evaluating the model on larger and more diverse datasets would provide a more comprehensive assessment of its robustness and generalization capabilities. Consequently, the collection of larger-scale datasets constitutes an important direction for future work.

Another promising research direction is the evaluation of qCNN on traffic collected from a broader range of VC platforms and communication environments. Such experiments would enable a more extensive analysis of the model's ability to generalize across applications and network conditions. Furthermore, the present study employed a single CNN HEAD architecture, namely ResNet-18, primarily due to computational constraints. Future work may investigate deeper ResNet variants and more recent convolutional architectures, such as ResNeXt and EfficientNet, to determine whether additional performance gains can be achieved.

Finally, recent advances in transformer-based architectures have demonstrated their effectiveness in modeling sequential and time-series data. Integrating transformer components into the proposed framework represents a particularly promising avenue for future research, potentially enabling richer temporal representations, improved prediction accuracy, and a deeper understanding of the latent processes governing QoE formation in encrypted communication systems.

%% file: sections/appendix.tex
\clearpage
\appendix

\section{System Requirements}
The proposed qCNN architecture has relatively low computational requirements and can be trained and deployed on consumer-grade hardware. All experiments reported in this study were conducted on a single laptop equipped with a dedicated GPU.
\begin{table}[htbp!]
    \begin{center}
        \caption{Computational environment parameters}
        \label{tab:specs}
        \begin{adjustbox}{max width=\textwidth}
            \begin{tabular}{| c | c | c |}
                \hline
                    Parameter & Model           & Specification  \\
                \hline
                    CPU       & 12th Gen Intel® Core™ i7-12700H × 20   & 14 cores (6 performance + 8 efficient), 20 threads \\ 
                \hline
                    RAM       & Kingston FURY Impact             & DDR4 CL20, 3200 Mz, 64 GB (32 + 32)\\ 
                \hline
                    GPU       & NVIDIA GeForce RTX 4060 Laptop GPU & 8 GB \\
                \hline
            \end{tabular}
        \end{adjustbox}
    \end{center}
\end{table}

\clearpage

\section{True versus Predicted Scatter Plots of Different Models}
\begin{figure}[htb!]
    \begin{tabular}{ccc}
        RF \cite{ho1995random} & SVM \cite{cortes1995support} & AdaBoost \cite{freund1995desicion} \\
        \includegraphics[width=0.33\linewidth]{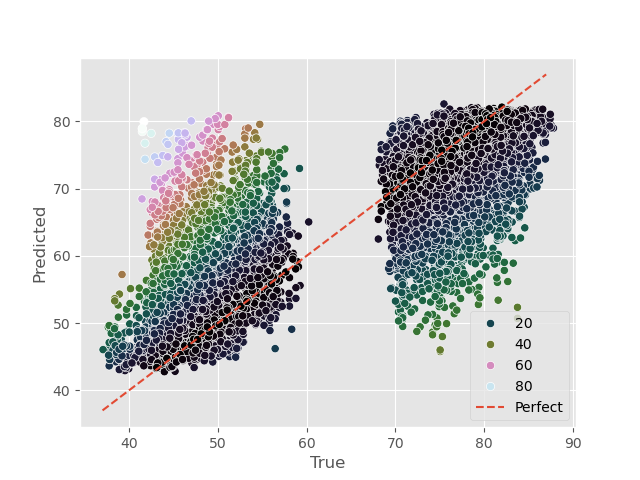} &
        \includegraphics[width=0.33\linewidth]{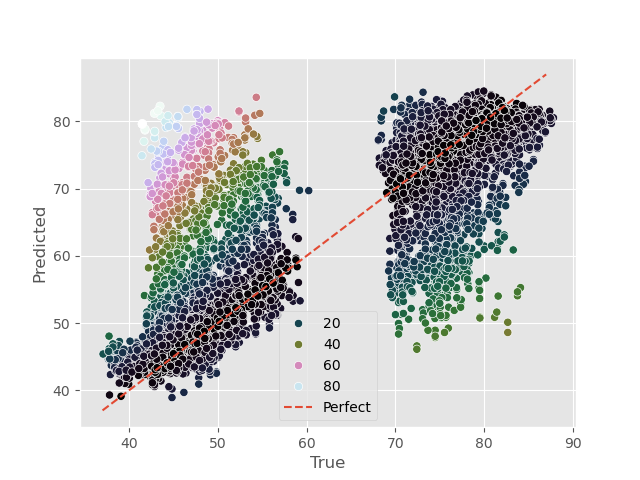} &
        \includegraphics[width=0.33\linewidth]{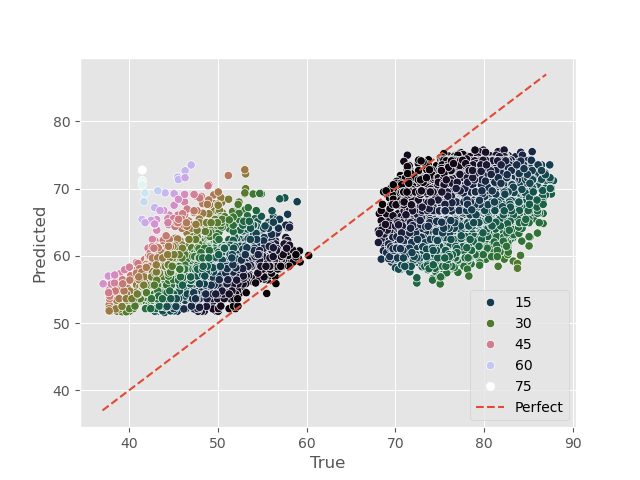} \\
        XGBoost \cite{chen2016xgboost} & CatBoost \cite{prokhorenkova2018catboost} & CNNQoE \cite{duc2020convolutional} \\
        \includegraphics[width=0.33\linewidth]{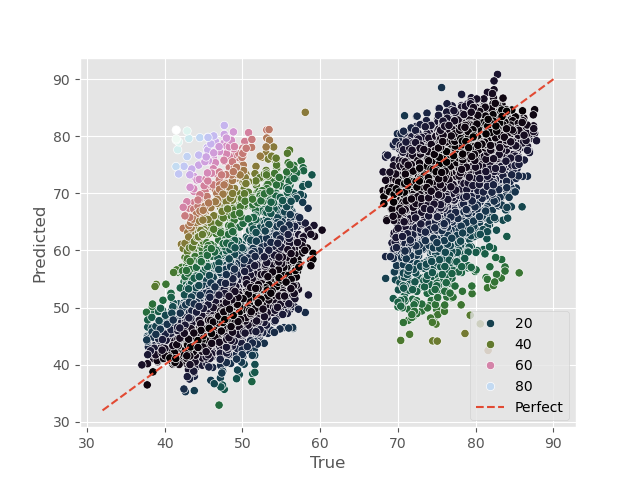} &
        \includegraphics[width=0.33\linewidth]{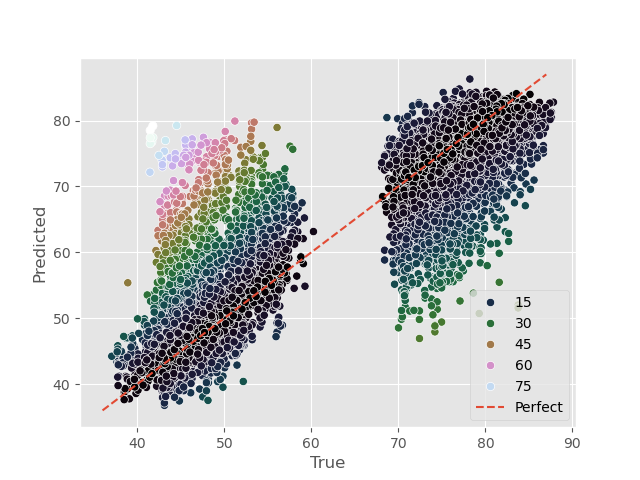} &
        \includegraphics[width=0.33\linewidth]{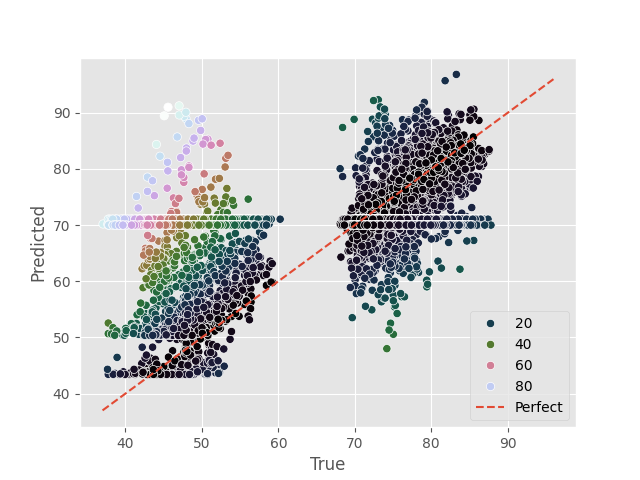} \\
        TCN \cite{bai2018empirical} & LSTM-QoE \cite{eswara2019streaming} & qCNN (Proposed)\\
        \includegraphics[width=0.33\linewidth]{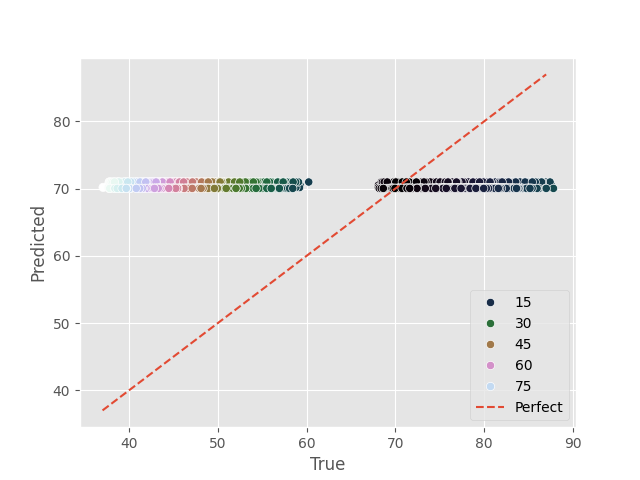} &
        \includegraphics[width=0.33\linewidth]{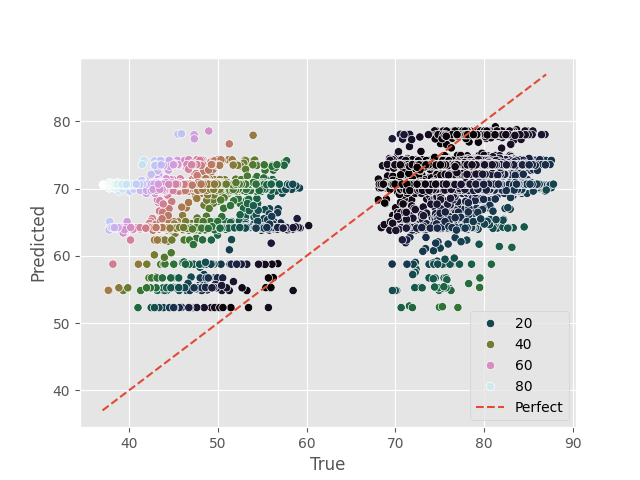} &
        \includegraphics[width=0.33\linewidth]{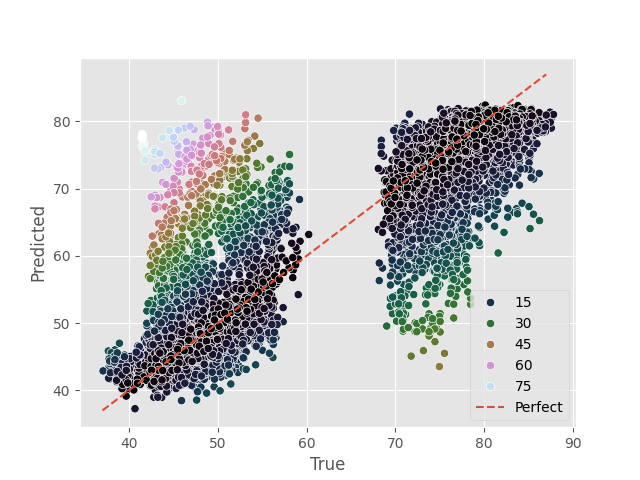} \\
    \end{tabular}
    \caption{Scatter plots comparing the predicted and ground-truth BRISQUE values for the evaluated models on the WA dataset.}
    \label{appdx-fig:wa-brisque-error-scatter}
\end{figure}

\clearpage

\begin{figure}[htb!]
    \begin{tabular}{ccc}
        RF \cite{ho1995random} & SVM \cite{cortes1995support} & AdaBoost \cite{freund1995desicion} \\
        \includegraphics[width=0.33\linewidth]{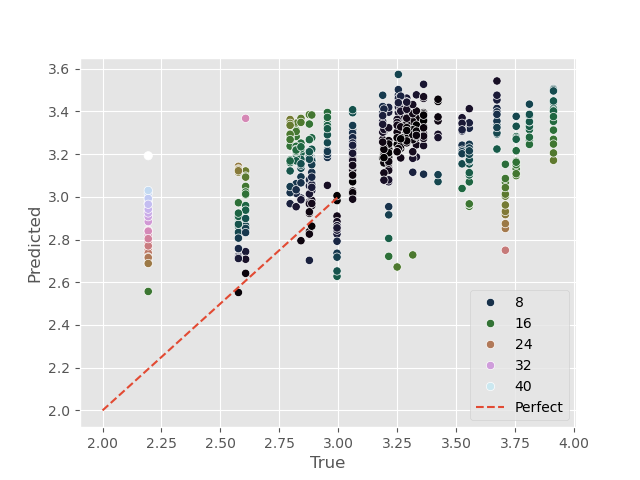} &
        \includegraphics[width=0.33\linewidth]{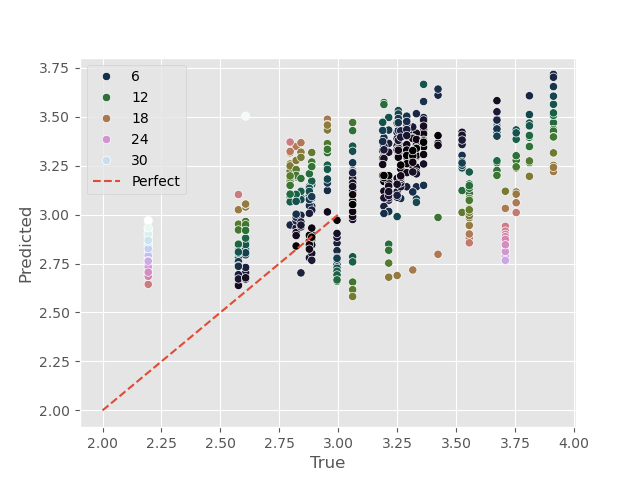} &
        \includegraphics[width=0.33\linewidth]{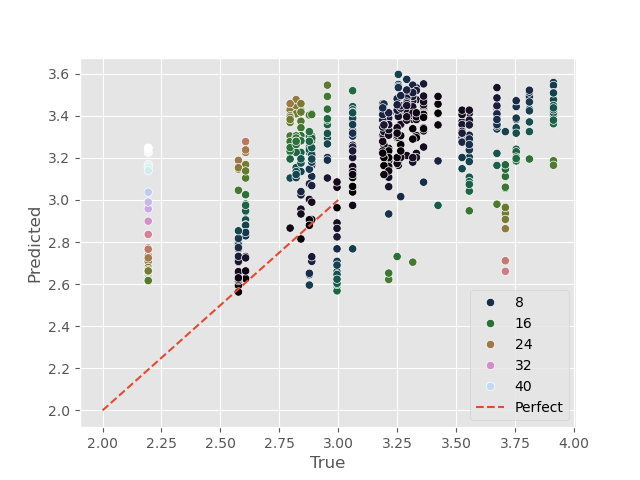} \\
        XGBoost \cite{chen2016xgboost} & CatBoost \cite{prokhorenkova2018catboost} & CNNQoE \cite{duc2020convolutional} \\
        \includegraphics[width=0.33\linewidth]{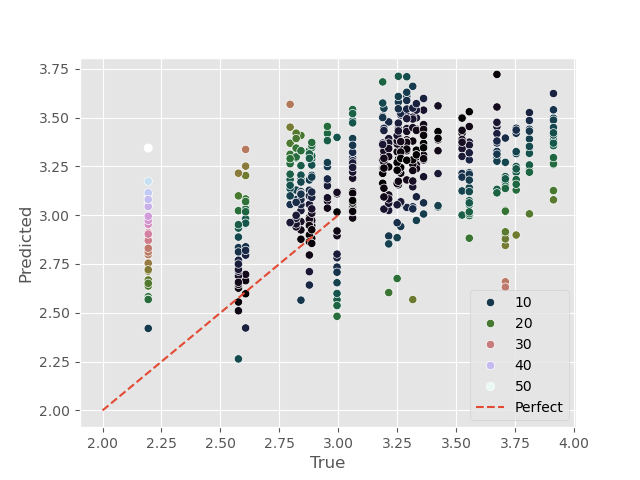} &
        \includegraphics[width=0.33\linewidth]{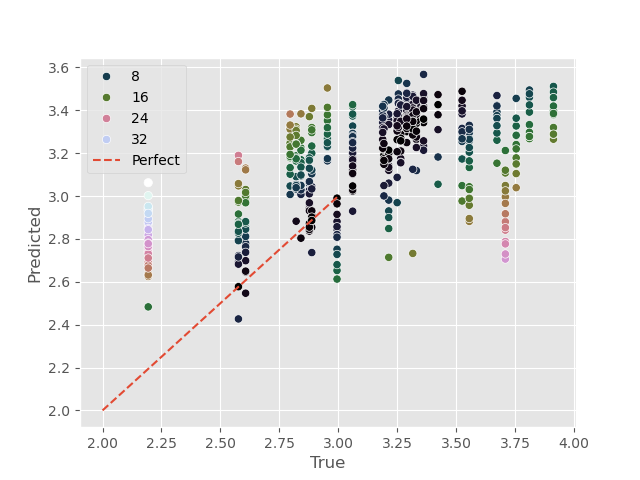} &
        \includegraphics[width=0.33\linewidth]{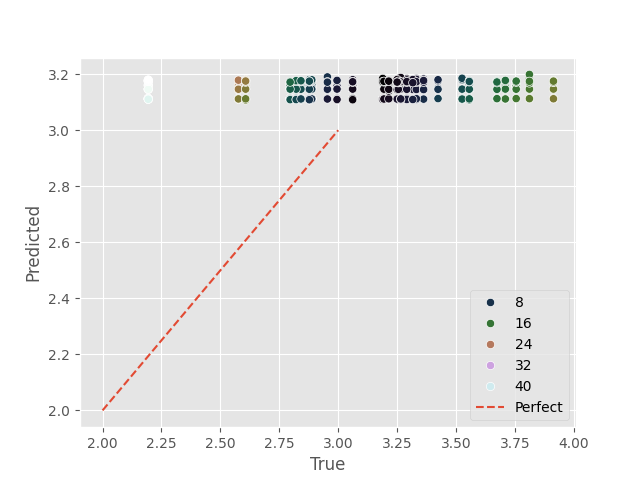} \\
        TCN \cite{bai2018empirical} & LSTM-QoE \cite{eswara2019streaming} & qCNN (Proposed)\\
        \includegraphics[width=0.33\linewidth]{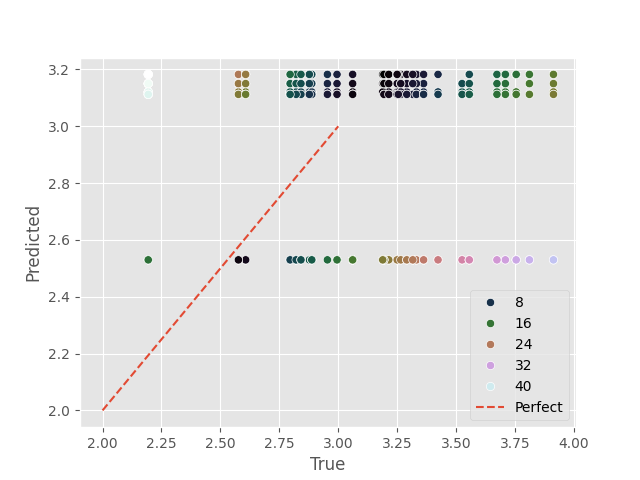} &
        \includegraphics[width=0.33\linewidth]{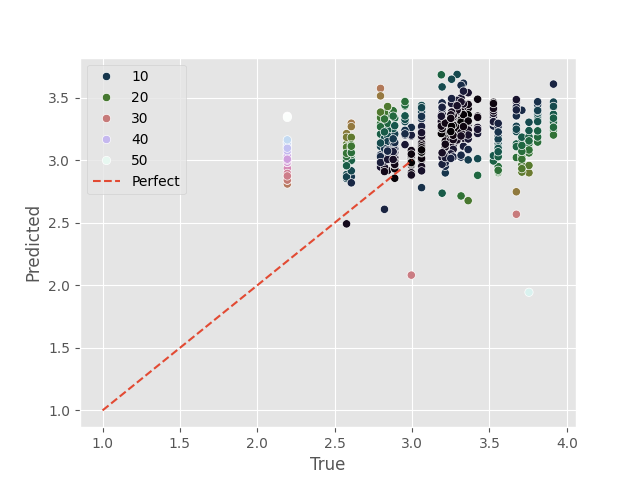} &
        \includegraphics[width=0.33\linewidth]{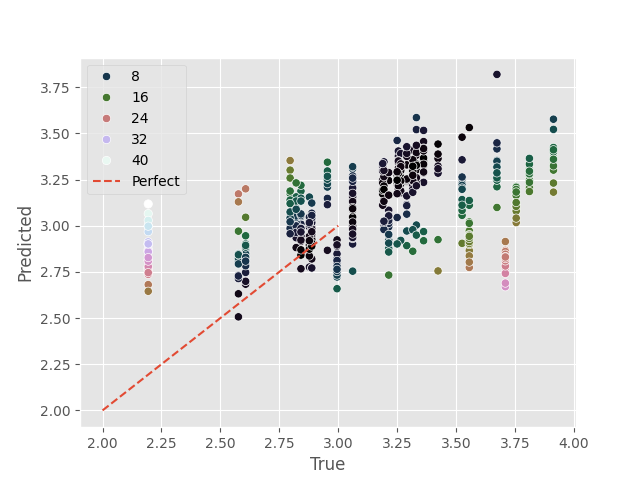} \\
    \end{tabular}
    \caption{Scatter plots comparing the predicted and ground-truth BRISQUE values for the evaluated models on the ZO dataset.}
    \label{appdx-fig:zo-mos-error-scatter}
\end{figure}

%% file: references.bib
@article{sidorov2024revisiting,
  title={Revisiting information cascades in online social networks},
  author={Sidorov, Michael and Hadar, Ofer and Vilenchik, Dan},
  journal={Mathematics},
  volume={13},
  number={1},
  pages={77},
  year={2024},
  publisher={MDPI}
}

@inproceedings{kornblith2019better,
  title={Do better imagenet models transfer better?},
  author={Kornblith, Simon and Shlens, Jonathon and Le, Quoc V},
  booktitle={Proceedings of the IEEE/CVF conference on computer vision and pattern recognition},
  pages={2661--2671},
  year={2019}
}

@inproceedings{he2016deep,
  title={Deep residual learning for image recognition},
  author={He, Kaiming and Zhang, Xiangyu and Ren, Shaoqing and Sun, Jian},
  booktitle={Proceedings of the IEEE conference on computer vision and pattern recognition},
  pages={770--778},
  year={2016}
}

@article{ebbehoj2022transfer,
  title={Transfer learning for non-image data in clinical research: A scoping review},
  author={Ebbehoj, Andreas and Thunbo, Mette {\O}stergaard and Andersen, Ole Emil and Glindtvad, Michala Vilstrup and Hulman, Adam},
  journal={PLOS Digital Health},
  volume={1},
  number={2},
  pages={e0000014},
  year={2022},
  publisher={Public Library of Science San Francisco, CA USA}
}

@article{sidorov2025estimating,
  title={Estimating QoE from Encrypted Video Conferencing Traffic},
  author={Sidorov, Michael and Birman, Raz and Hadar, Ofer and Dvir, Amit},
  journal={Sensors (Basel, Switzerland)},
  volume={25},
  number={4},
  pages={1009},
  year={2025}
}

@article{sidorov2025reliable,
  title={Reliable QoE Prediction in IMVCAs Using an LMM-Based Agent},
  author={Sidorov, Michael and Berger, Tamir and Sterenson, Jonathan and Birman, Raz and Hadar, Ofer},
  journal={Sensors},
  volume={25},
  number={14},
  pages={4450},
  year={2025},
  publisher={MDPI}
}

@inproceedings{wassermann2019see,
  title={I see what you see: Real time prediction of video quality from encrypted streaming traffic},
  author={Wassermann, Sarah and Seufert, Michael and Casas, Pedro and Gang, Li and Li, Kuang},
  booktitle={Proceedings of the 4th Internet-QoE Workshop on QoE-based Analysis and Management of Data Communication Networks},
  pages={1--6},
  year={2019}
}

@article{torres2017unsupervised,
  title={Unsupervised deep learning for real-time assessment of video streaming services},
  author={Torres Vega, Maria and Mocanu, Decebal Constantin and Liotta, Antonio},
  journal={Multimedia Tools and Applications},
  volume={76},
  number={21},
  pages={22303--22327},
  year={2017},
  publisher={Springer}
}

@article{raca2020leveraging,
  title={On leveraging machine and deep learning for throughput prediction in cellular networks: Design, performance, and challenges},
  author={Raca, Darijo and Zahran, Ahmed H and Sreenan, Cormac J and Sinha, Rakesh K and Halepovic, Emir and Jana, Rittwik and Gopalakrishnan, Vijay},
  journal={IEEE Communications Magazine},
  volume={58},
  number={3},
  pages={11--17},
  year={2020},
  publisher={IEEE}
}

@inproceedings{bentaleb2019bandwidth,
  title={Bandwidth prediction in low-latency chunked streaming},
  author={Bentaleb, Abdelhak and Timmerer, Christian and Begen, Ali C and Zimmermann, Roger},
  booktitle={Proceedings of the 29th ACM workshop on network and operating systems support for digital audio and video},
  pages={7--13},
  year={2019}
}

@inproceedings{mao2017neural,
  title={Neural adaptive video streaming with pensieve},
  author={Mao, Hongzi and Netravali, Ravi and Alizadeh, Mohammad},
  booktitle={Proceedings of the conference of the ACM special interest group on data communication},
  pages={197--210},
  year={2017}
}

@article{hinton2006reducing,
  title={Reducing the dimensionality of data with neural networks},
  author={Hinton, Geoffrey E and Salakhutdinov, Ruslan R},
  journal={science},
  volume={313},
  number={5786},
  pages={504--507},
  year={2006},
  publisher={American Association for the Advancement of Science}
}

@article{kingma2014adam,
  title={Adam: A method for stochastic optimization},
  author={Kingma, Diederik P and Ba, Jimmy},
  journal={arXiv preprint arXiv:1412.6980},
  year={2014}
}

@article{srivastava2014dropout,
  title={Dropout: a simple way to prevent neural networks from overfitting},
  author={Srivastava, Nitish and Hinton, Geoffrey and Krizhevsky, Alex and Sutskever, Ilya and Salakhutdinov, Ruslan},
  journal={The journal of machine learning research},
  volume={15},
  number={1},
  pages={1929--1958},
  year={2014},
  publisher={JMLR. org}
}

@article{dickey1979distribution,
  title={Distribution of the estimators for autoregressive time series with a unit root},
  author={Dickey, David A and Fuller, Wayne A},
  journal={Journal of the American statistical association},
  volume={74},
  number={366a},
  pages={427--431},
  year={1979},
  publisher={Taylor \& Francis}
}

@inproceedings{smith2017cyclical,
  title={Cyclical learning rates for training neural networks},
  author={Smith, Leslie N},
  booktitle={2017 IEEE winter conference on applications of computer vision (WACV)},
  pages={464--472},
  year={2017},
  organization={IEEE}
}

@article{feamster2017and,
  title={Why (and how) networks should run themselves},
  author={Feamster, Nick and Rexford, Jennifer},
  journal={arXiv preprint arXiv:1710.11583},
  year={2017}
}

@inproceedings{naylor2014cost,
  title={The cost of the" s" in https},
  author={Naylor, David and Finamore, Alessandro and Leontiadis, Ilias and Grunenberger, Yan and Mellia, Marco and Munaf{\`o}, Maurizio and Papagiannaki, Konstantina and Steenkiste, Peter},
  booktitle={Proceedings of the 10th ACM International on Conference on emerging Networking Experiments and Technologies},
  pages={133--140},
  year={2014}
}

@article{segal2023using,
  title={Using EfficientNet-B7 (CNN), variational auto encoder (VAE) and Siamese twins’ networks to evaluate human exercises as super objects in a TSSCI images},
  author={Segal, Yoram and Hadar, Ofer and Lhotska, Lenka},
  journal={Journal of personalized medicine},
  volume={13},
  number={5},
  pages={874},
  year={2023},
  publisher={MDPI}
}

@inproceedings{shapira2019flowpic,
  title={Flowpic: Encrypted internet traffic classification is as easy as image recognition},
  author={Shapira, Tal and Shavitt, Yuval},
  booktitle={IEEE INFOCOM 2019-IEEE conference on computer communications workshops (INFOCOM WKSHPS)},
  pages={680--687},
  year={2019},
  organization={IEEE}
}

@article{cortes1995support,
  title={Support-vector networks},
  author={Cortes, Corinna and Vapnik, Vladimir},
  journal={Machine learning},
  volume={20},
  number={3},
  pages={273--297},
  year={1995},
  publisher={Springer}
}

@inproceedings{freund1995desicion,
  title={A desicion-theoretic generalization of on-line learning and an application to boosting},
  author={Freund, Yoav and Schapire, Robert E},
  booktitle={European conference on computational learning theory},
  pages={23--37},
  year={1995},
  organization={Springer}
}

@inproceedings{ho1995random,
  title={Random decision forests},
  author={Ho, Tin Kam},
  booktitle={Proceedings of 3rd international conference on document analysis and recognition},
  volume={1},
  pages={278--282},
  year={1995},
  organization={IEEE}
}

@article{lopez2018deep,
  title={Deep learning model for multimedia quality of experience prediction based on network flow packets},
  author={Lopez-Martin, Manuel and Carro, Belen and Lloret, Jaime and Egea, Santiago and Sanchez-Esguevillas, Antonio},
  journal={IEEE Communications Magazine},
  volume={56},
  number={9},
  pages={110--117},
  year={2018},
  publisher={IEEE}
}

@article{duc2020convolutional,
  title={Convolutional neural networks for continuous QoE prediction in video streaming services},
  author={Duc, Tho Nguyen and Minh, Chanh Tran and Xuan, Tan Phan and Kamioka, Eiji},
  journal={IEEE Access},
  volume={8},
  pages={116268--116278},
  year={2020},
  publisher={IEEE}
}

@article{bai2018empirical,
  title={An empirical evaluation of generic convolutional and recurrent networks for sequence modeling},
  author={Bai, Shaojie and Kolter, J Zico and Koltun, Vladlen},
  journal={arXiv preprint arXiv:1803.01271},
  year={2018}
}

@article{eswara2019streaming,
  title={Streaming video QoE modeling and prediction: A long short-term memory approach},
  author={Eswara, Nagabhushan and Ashique, S and Panchbhai, Anand and Chakraborty, Soumen and Sethuram, Hemanth P and Kuchi, Kiran and Kumar, Abhinav and Channappayya, Sumohana S},
  journal={IEEE Transactions on Circuits and Systems for Video Technology},
  volume={30},
  number={3},
  pages={661--673},
  year={2019},
  publisher={IEEE}
}

@article{prokhorenkova2018catboost,
  title={CatBoost: unbiased boosting with categorical features},
  author={Prokhorenkova, Liudmila and Gusev, Gleb and Vorobev, Aleksandr and Dorogush, Anna Veronika and Gulin, Andrey},
  journal={Advances in neural information processing systems},
  volume={31},
  year={2018}
}

@inproceedings{chen2016xgboost,
  title={Xgboost: A scalable tree boosting system},
  author={Chen, Tianqi and Guestrin, Carlos},
  booktitle={Proceedings of the 22nd acm sigkdd international conference on knowledge discovery and data mining},
  pages={785--794},
  year={2016}
}

@inproceedings{mittal2011blind,
  title={Blind/referenceless image spatial quality evaluator},
  author={Mittal, Anish and Moorthy, Anush K and Bovik, Alan C},
  booktitle={2011 conference record of the forty fifth asilomar conference on signals, systems and computers (ASILOMAR)},
  pages={723--727},
  year={2011},
  organization={IEEE}
}

@misc{recommendation20081080,
  title={1080-Quality of experience requirements for IPTV services},
  author={Recommendation, ITUTG},
  year={2008},
  publisher={ITU}
}

@inproceedings{dimopoulos2016measuring,
  title={Measuring video QoE from encrypted traffic},
  author={Dimopoulos, Giorgos and Leontiadis, Ilias and Barlet-Ros, Pere and Papagiannaki, Konstantina},
  booktitle={Proceedings of the 2016 Internet Measurement Conference},
  pages={513--526},
  year={2016}
}

@article{orsolic2017machine,
  title={A machine learning approach to classifying YouTube QoE based on encrypted network traffic},
  author={Orsolic, Irena and Pevec, Dario and Suznjevic, Mirko and Skorin-Kapov, Lea},
  journal={Multimedia tools and applications},
  volume={76},
  number={21},
  pages={22267--22301},
  year={2017},
  publisher={Springer}
}

@article{bronzino2019inferring,
  title={Inferring streaming video quality from encrypted traffic: Practical models and deployment experience},
  author={Bronzino, Francesco and Schmitt, Paul and Ayoubi, Sara and Martins, Guilherme and Teixeira, Renata and Feamster, Nick},
  journal={Proceedings of the ACM on Measurement and Analysis of Computing Systems},
  volume={3},
  number={3},
  pages={1--25},
  year={2019},
  publisher={ACM New York, NY, USA}
}

@inproceedings{gutterman2019requet,
  title={Requet: Real-time QoE detection for encrypted YouTube traffic},
  author={Gutterman, Craig and Guo, Katherine and Arora, Sarthak and Wang, Xiaoyang and Wu, Les and Katz-Bassett, Ethan and Zussman, Gil},
  booktitle={Proceedings of the 10th ACM Multimedia Systems Conference},
  pages={48--59},
  year={2019}
}

@inproceedings{oura2024qoe,
  title={Qoe estimation method with time-series features extracted from packet flows for video streaming},
  author={Oura, Junki and Wada, Akihiro and Ogawa, Masatoshi and Yokoo, Kaoru and Kakuta, Jun and Ninomiya, Teruhisa},
  booktitle={2024 IEEE 21st Consumer Communications \& Networking Conference (CCNC)},
  pages={1--6},
  year={2024},
  organization={IEEE}
}

@inproceedings{shen2020deepqoe,
  title={DeepQoE: Real-time measurement of video QoE from encrypted traffic with deep learning},
  author={Shen, Meng and Zhang, Jinpeng and Xu, Ke and Zhu, Liehuang and Liu, Jiangchuan and Du, Xiaojiang},
  booktitle={2020 IEEE/ACM 28th International Symposium on Quality of Service (IWQoS)},
  pages={1--10},
  year={2020},
  organization={IEEE}
}

@inproceedings{panchenko2016website,
  title={Website Fingerprinting at Internet Scale.},
  author={Panchenko, Andriy and Lanze, Fabian and Pennekamp, Jan and Engel, Thomas and Zinnen, Andreas and Henze, Martin and Wehrle, Klaus},
  booktitle={NDSS},
  volume={1},
  pages={23477},
  year={2016}
}

@inproceedings{chen2017seq2img,
  title={Seq2img: A sequence-to-image based approach towards ip traffic classification using convolutional neural networks},
  author={Chen, Zhitang and He, Ke and Li, Jian and Geng, Yanhui},
  booktitle={2017 IEEE International conference on big data (big data)},
  pages={1271--1276},
  year={2017},
  organization={IEEE}
}

@article{rezaei2018deep,
  title={Deep learning for encrypted traffic classification: An overview},
  author={Rezaei, Shahbaz and Liu, Xin},
  journal={arXiv preprint arXiv:1810.07906},
  year={2018}
}

@inproceedings{aceto2018mobile,
  title={Mobile encrypted traffic classification using deep learning},
  author={Aceto, Giuseppe and Ciuonzo, Domenico and Montieri, Antonio and Pescap{\'e}, Antonio},
  booktitle={2018 Network traffic measurement and analysis conference (TMA)},
  pages={1--8},
  year={2018},
  organization={IEEE}
}

@inproceedings{vu2017deep,
  title={A deep learning based method for handling imbalanced problem in network traffic classification},
  author={Vu, Ly and Bui, Cong Thanh and Nguyen, Quang Uy},
  booktitle={Proceedings of the 8th international symposium on information and communication technology},
  pages={333--339},
  year={2017}
}

@article{rezaei2018achieve,
  title={How to achieve high classification accuracy with just a few labels: A semi-supervised approach using sampled packets},
  author={Rezaei, Shahbaz and Liu, Xin},
  journal={arXiv preprint arXiv:1812.09761},
  year={2018}
}

@article{lotfollahi2020deep,
  title={Deep packet: A novel approach for encrypted traffic classification using deep learning},
  author={Lotfollahi, Mohammad and Jafari Siavoshani, Mahdi and Shirali Hossein Zade, Ramin and Saberian, Mohammdsadegh},
  journal={Soft Computing},
  volume={24},
  number={3},
  pages={1999--2012},
  year={2020},
  publisher={Springer}
}

@article{lopez2017network,
  title={Network traffic classifier with convolutional and recurrent neural networks for Internet of Things},
  author={Lopez-Martin, Manuel and Carro, Belen and Sanchez-Esguevillas, Antonio and Lloret, Jaime},
  journal={IEEE access},
  volume={5},
  pages={18042--18050},
  year={2017},
  publisher={IEEE}
}

@inproceedings{hochst2017unsupervised,
  title={Unsupervised traffic flow classification using a neural autoencoder},
  author={H{\"o}chst, Jonas and Baumg{\"a}rtner, Lars and Hollick, Matthias and Freisleben, Bernd},
  booktitle={2017 IEEE 42Nd Conference on local computer networks (LCN)},
  pages={523--526},
  year={2017},
  organization={IEEE}
}
